\documentclass[altaffilletter,aps,nofootinbib,twocolumn,prd,eqsecnum,preprintnumbers,superscriptaddress,10pt,floatfix]{revtex4-2}
\pdfoutput=1
\usepackage{amsmath}
\usepackage{amssymb}
\usepackage{bm}
\usepackage{subfigure}
\usepackage{graphicx}
\usepackage{dcolumn}
\usepackage{booktabs}
\usepackage{amsfonts}
\usepackage{mathtools}
\usepackage{enumerate}
\usepackage{orcidlink}
\usepackage{xcolor}
\usepackage{mathrsfs}
\usepackage{epstopdf}
\usepackage{url}
\usepackage{footnote}
\usepackage{textcomp}
\usepackage{dsfont}
\usepackage{ulem}
\usepackage{enumerate}   
\usepackage{comment}
\usepackage{appendix}
\usepackage{textcomp}
\usepackage{tipa}
\hypersetup{
	colorlinks=true,
	linkcolor=red!70!black,     % \eqref{}, \ref{}, internal links
	citecolor=cyan!99!black,    % \cite{}
	urlcolor=blue!55!black      % URLs
}

\renewcommand{\boxed}[1]{#1}
\newcommand{\dd}{\mathrm{d}}
\newcommand{\e}{\mathrm{e}}
\newcommand{\Mbare}{M_{\rm Sch}}
\newcommand{\rhSch}{r_h^{\rm Sch}}
\newcommand{\rphSch}{r_{\rm ph}^{\rm Sch}}
\newcommand{\bphSch}{b_{\rm ph}^{\rm Sch}}
\newcommand{\Mh}{M_{\mathrm{H}}}
\newcommand{\Meff}{M_{\bullet,\gamma}}
\newcommand{\Mext}{M_{\mathrm{ext},\gamma}}
\newcommand{\rin}{r_{\mathrm{in},\gamma}}
\newcommand{\rh}{r_{h,\gamma}}

\makeatletter
\newcommand*{\rom}[1]{\expandafter\@slowromancap\romannumeral #1@}
\makeatother

\begin{document}

\title{
Black holes in depleted Dehnen dark matter halos with physical inner edges and strong-field observables
}

\author{Hassan Hassanabadi\,\orcidlink{0000-0001-7487-6898}
}
\email{hassanhassanabadi@mail.fresnostate.edu}
\affiliation{Physics Department, California State University, Fresno, CA 93740, USA}

\author{Soroush Zare\,\orcidlink{0000-0003-0748-3386}
}
\email{soroush.z.zare@helsinki.fi}
\affiliation{ Helsinki Institute of Physics, University of Helsinki, P.O. Box 64, FI-00014 Helsinki, Finland}

\begin{abstract}
We construct a static and spherically symmetric black hole (BH) spacetime surrounded by a depleted Dehnen dark matter (DM) halo.  The model describes a final equilibrium configuration rather than dynamical accretion process: the initial halo mass inside a prescribed inner edge is incorporated into the central BH, leaving a vacuum gap between the dressed horizon and the surviving halo. Consequently, the horizon radius depends on the Dehnen cusp index $\gamma$ and the absorbed halo mass, while the Schwarzschild result is recovered exactly when the halo is removed.  We derive general expressions for the metric functions and redshift phase and obtain explicit solutions for $\gamma=0,1,2,$ and $5/2$.  The dominant energy condition imposes the universal local bound $\kappa\geq5/4$, whereas global admissibility depends additionally on the cusp index and halo compactness.  A literal stable Einstein-cluster interpretation requires the stronger condition $\kappa\geq3$ when stable circular motion is imposed at the halo edge.  We also analyze the photon sphere, shadow, and admissible parameter space, demonstrating how the halo cusp and depletion scale affect strong-field observables.  Previously studied analytic BH--halo geometries and the Schwarzschild and weak-halo limits follow as special cases of the general construction.

\end{abstract}

\maketitle

%%%%%%%%%%%%%%%
%%%%%%%%%%%%%%%
%%%%%%%%%%%%%%%

\section{Introduction}\label{sec.intro}

\label{sec:introduction}

Supermassive BHs are not isolated objects, but reside in galactic nuclei containing dense stellar systems, gaseous structures, and potentially cusped or spiked DM distributions \cite{Genzel2010,GondoloSilk1999,Sadeghian2013}.  Nevertheless, vacuum BH geometries provide the natural zeroth-order description of the strong-field region because, for ordinary astrophysical environments, the environmental stress-energy is typically perturbatively small compared with the characteristic curvature scale generated by the BH \cite{Barausse2014}.  Precision observations make corrections to this approximation increasingly relevant.  Environmental matter can alter orbital frequencies, gravitational redshift, lensing, shadows, and the phase and propagation of gravitational waves, and it can therefore imitate or bias small departures from an isolated BH \cite{Barausse2014,Cardoso2022,Figueiredo2023}.  The central theoretical problem is consequently not only to prescribe a galactic density, but to embed it in a relativistic geometry whose horizon, mass budget, matching conditions, and matter interpretation are mutually consistent.

The distribution near a BH need not be the inward extrapolation of the halo measured on galactic scales.  Adiabatic BH growth may steepen an initial cusp into a DM spike \cite{GondoloSilk1999,Quinlan1995}, whereas mergers, stellar scattering, self-annihilation, and capture by the hole can soften, truncate, or deplete that spike \cite{Merritt2002,Sadeghian2013}. Relativistic calculations of the phase-space distribution make the capture boundary explicit and show that the density is suppressed in the innermost region \cite{Sadeghian2013}.  
In binaries, the orbiting secondary can itself modify the surrounding DM distribution, so a permanently fixed spike is not always a self-consistent approximation
 \cite{Kavanagh2020,Coogan2022}. These mechanisms do not predict a universal vacuum gap with a unique inner edge. Rather, they motivate treating the inner support of a static halo as a physical input to be constrained, instead of extending a phenomenological galactic profile unchanged through the horizon.

For that purpose the Dehnen family is especially useful.  Its spherical density behaves as $\rho\sim r^{-\gamma}$ at small radius and as $\rho\sim r^{-4}$ at large radius, giving a finite total mass while allowing the central logarithmic slope $0\leq\gamma<3$ to be varied independently of the halo mass and scale radius \cite{Dehnen1993}.  It contains a cored model at $\gamma=0$, the Hernquist profile at $\gamma=1$ \cite{Hernquist1990}, the Jaffe model at $\gamma=2$ \cite{Jaffe1983}, and the steeper $\gamma=5/2$ case considered below.  The family therefore provides a controlled way to ask which strong-field properties follow from the total halo mass and which retain information about the unresolved central cusp.

Several relativistic constructions have placed BHs inside anisotropic matter distributions.  The Einstein-cluster picture represents collisionless particles on statistically isotropic circular orbits; it has vanishing radial pressure but nonzero tangential pressure \cite{Einstein1939,GeralicoEtAl2012}.  Modern realizations have produced asymptotically flat BH geometries for specific or generic halo profiles and examined their geodesics, shadows, and gravitational-wave signatures \cite{Cardoso2022,PezzellaEtAl2025,Jusufi2022,Konoplya2022,Figueiredo2023,Acharyya2024}.  In particular, Cardoso \emph{et al.} obtained an analytic BH--Hernquist solution with a regular horizon \cite{Cardoso2022}, while Shen, Wang, and Yin introduced an independent inner halo radius and used the energy conditions to constrain it \cite{Shen2025}.  Maeda, Cardoso, and Wang subsequently constructed central spiky Einstein clusters, imposed an inner matter edge at a marginally stable circular orbit, and exhibited both finite-radius and infinite-radius matter distributions \cite{Maeda2025}.  
%These studies establish that relativistic halo models can be constructed consistently, but they begin from different prescriptions for the central mass, the horizon, and the inner support of the matter.  It is therefore not immediate how the mass removed from a common galactic seed changes the BH itself, or how that change depends on the seed cusp.  This issue requires a framework in which the event horizon and the inner edge of the surviving matter are treated as distinct surfaces and the corresponding mass budget is explicit.
These studies demonstrate that consistent relativistic BH--halo geometries can be constructed, but they adopt different prescriptions for the central mass, the horizon, and the inner support of the matter. Consequently, they do not directly determine how depletion of a common galactic seed modifies the central BH, or how the resulting change depends on the seed cusp. Addressing this question calls for a framework in which the event horizon and the inner edge of the surviving halo are treated as distinct surfaces, with the associated mass budget made explicit.

This distinction motivates the present depleted-halo construction.  A literal static Einstein cluster cannot extend to the horizon because massive particles cannot remain on circular timelike orbits there.  Moreover, DM captured during the formation or relaxation of the central object should contribute to the dressed BH mass rather than remain simultaneously counted in the exterior halo.  
We therefore formulate a static final state, not a time-dependent collapse or accretion solution.  Starting from a Dehnen seed, we assign the mass initially contained inside a prescribed edge to the central object, determine the resulting cusp-dependent horizon, and normalize the surviving exterior halo to the remaining mass.  The vacuum and matter regions are then matched at the halo edge, and the absence of an unphysical surface layer and of additional exterior horizons is examined explicitly. Within the same construction we derive the general metric, constrain the gap using the energy and circular-orbit conditions, and determine whether the photon sphere lies in the vacuum gap or in the halo.  The analytic geometries of Refs.~\cite{Shen2025,Cardoso2022} arise from specified limits of this common framework.

The paper is organized as follows.  
Section~\ref{sec:construction} distinguishes the dressed horizon from the halo edge and sets out the final-state construction. Sections~\ref{sec:horizon} and \ref{sec:density} determine the dressed horizon and the depleted exterior mass distribution.  The general metric is derived in Sec.~\ref{sec:metric}, followed by the four explicit Dehnen cases in Sec.~\ref{sec:dehnen-cases}.  Energy conditions, their numerical verification, and orbital stability are analyzed in Sec.~\ref{sec:admissibility}; photon spheres and shadow scales are studied in Sec.~\ref{sec:photon}, including their angular frequencies and instability timescales.  The corresponding fixed-final-mass limits of Refs.~\cite{Shen2025,Cardoso2022} are identified within the $\gamma=1$ solution.  Section~\ref{sec:conclusion} summarizes the physical conclusions. Throughout this work we use geometrized units, $G=c=1$.

%%%%%%%%%%%%%%%
%%%%%%%%%%%%%%%
%%%%%%%%%%%%%%%

%\section{Horizon and halo-edge radii}\label{sec:radii}
\section{Physical construction, vacuum gap, and scope}
\label{sec:construction}

We first separate the geometric horizon from the matter boundary and then
relate our edge parameter to the notation of Ref.~\cite{Shen2025}.  
This distinction is the starting point for the mass bookkeeping used below.

In the density profile of Ref.~\cite{Shen2025}, the parameter
$\nu_{\rm SWY}$
fixes the inner edge of the halo,
$r_{\rm in}=\nu_{\rm SWY}M_{\rm BH}$.
%; it is not the event horizon coefficient.  The region below the halo edge is matched to a Schwarzschild geometry, whose horizon is $2M_{\rm BH}$.  Replacing $\nu_{\rm SWY}$ by an arbitrary power such as $\nu_{\rm SWY}^{1+\gamma}$ therefore changes only the vacuum gap.  If the same replacement is inserted into the vacuum lapse, it merely redefines the BH mass and incorrectly leaves a $\gamma$ dependence when the halo is absent.

Throughout our construction the independent edge parameter is $\kappa\equiv r_{\rm in}/r_h$.  
%The symbol $\nu_{\rm SWY}$ is used only when recovering the notation of Ref.~\cite{Shen2025}; the two conventions are related by $\nu_{\rm SWY}=2\kappa$ because their construction fixes $r_h=2M_{\rm BH}$.  We do not alternate between them elsewhere.

We instead distinguish a dressed horizon $\rh$ from the halo edge $\rin$. The model has three positive input scales, $\Mbare>0$, $\Mh>0$, and $a>0$. Here $\Mbare$ is the mass of the reference Schwarzschild geometry obtained when the halo is removed, $\Mh$ is the total mass of an initial Dehnen distribution, and $a$ is its scale radius.  Its standard reference scales are
\begin{eqnarray}
\nonumber
&&\rhSch=2\Mbare,\quad \rphSch=3\Mbare, \\
\nonumber
&&r_{\rm mb}^{\rm Sch}=4\Mbare,\quad r_{\rm ISCO}^{\rm Sch}=6\Mbare,\\
&&\bphSch=3\sqrt3\Mbare,\quad 
\Omega_{\rm LR}^{\rm Sch}=\Lambda_{\rm LR}^{\rm Sch}=\dfrac{1}{3\sqrt3\Mbare}.
\label{eq:schwarzschild_reference_scales}
\end{eqnarray}
All numerical radii and observables below are reported relative to these Schwarzschild values.    
The inner halo edge is parameterized as $\rin=\kappa\rh$, with $\kappa>1$, so that the matter distribution begins outside the event horizon.
 A stronger lower bound, $\kappa\geq 5/4$, follows from the dominant energy condition in a one-sided neighborhood of the halo edge, as derived below and in
agreement with Ref.~\cite{Shen2025}.
The exact global bound can be stronger and is then profile-dependent.  If one demands stable circular particle orbits all the way to the halo edge, the more conservative Schwarzschild-motivated choice is $\kappa\simeq3$.

%These definitions leave the physical origin of the gap open; the next section specifies the final-state interpretation adopted in this work.
These definitions leave the physical origin of the gap open. We next specify
the final-state interpretation underlying the present construction.

%%%%%%%%%%%%%%%
%%%%%%%%%%%%%%%
%%%%%%%%%%%%%%%

%\section{Physical construction, vacuum gap, and scope}
%\label{sec:construction}

We now state the physical assumptions, divide the spacetime into its interior, gap, and halo regions, and impose the matching conditions at the halo edge.

The metric derived below describes an idealized final static configuration rather than the dynamical formation of a BH in an otherwise unchanged galaxy.  
A possible history motivating the construction is the growth of a central seed BH accompanied by capture or scattering of the innermost DM particles. After relaxation, the matter that remains in the stationary halo has support only outside $\rin$.  Accordingly, the spacetime is divided into the three radial regions summarized in Table~ \ref{tab:regions}:
\begin{table*}[t]
	\centering
	\caption{Radial regions of the spacetime and their physical interpretation.}
	\label{tab:regions}
	\begin{tabular}{@{}lll@{}}
		\toprule
		Region & Radial range & Interpretation \\
		\midrule
		I   & $0<r\leq\rh$      & BH interior; no static halo interpretation \\
		II  & $\rh<r<\rin$      & Schwarzschild vacuum gap \\
		III & $r\geq\rin$        & Final stationary, anisotropic DM halo \\
		\bottomrule
	\end{tabular}
\end{table*}
The phrase ``vacuum gap'' means $T^{\mu}{}_{\nu} = \text{diag} (\rho_\gamma(r)=0, P_{r,\gamma}(r)=0, P_{t,\gamma}(r)=0, P_{t,\gamma}(r)=0)$ for $\rh<r<\rin$.
%\begin{eqnarray}
%&&T^{\mu}{}_{\nu}=0,\\
%&&\rho_\gamma(r)=P_{r,\gamma}(r)=P_{t,\gamma}(r)=0,\\
%&&\rh<r<\rin.
%	\label{eq:vacuum_gap_source}
%\end{eqnarray}
%It does not mean that DM can never fall through this region.  Rather, there is no stationary Einstein-cluster component there in the final model.  Matter that crosses the horizon ceases to be separately identifiable as exterior halo mass and contributes to the dressed BH mass.

The term ``vacuum gap'' should not be interpreted as excluding dynamical infall of DM through this region. Rather, it indicates that no stationary Einstein-cluster component is present there in the final configuration. Within this final-state interpretation, DM captured by the BH is no longer part of the exterior halo and instead contributes to the mass of the central BH~\cite{MachOdrzywolek2021,Shapiro2023,LoraClavijoEtAl2014}.

The most conservative statement is that the original inner mass has been removed from the final stationary halo, partly by capture and possibly partly by outward redistribution.  The present analytic model makes the stronger, maximal-accretion assumption that all of it is incorporated into the BH.

The metric functions $A_\gamma(r)$ and $B_\gamma(r)$ introduced in
Eq.~(5.1) determine, respectively, the temporal redshift and the radial
geometry. Since Region~II is vacuum, Birkhoff's theorem implies that the
geometry there is locally Schwarzschild, up to a constant normalization of
the timelike Killing coordinate~\cite{Poisson2004}. Hence
\begin{equation}
	B_{\rm gap}(r)=1-\frac{\rh}{r},
	\qquad
	A_{\rm gap}(r)=C_\gamma
	\left(1-\frac{\rh}{r}\right),
	\label{eq:gap_metric_early}
\end{equation}
where $C_\gamma>0$ is fixed by matching to the exterior halo at $r=\rin$.
This constant changes the redshift relative to infinity but cannot move the
horizon: the only zero in the gap remains $r=\rh$.  The smooth $n=1$ density
introduced in Sec.~\ref{sec:density} vanishes at the edge and makes both the
exterior mass and its first derivative vanish there.  Consequently the two
metric functions and their first derivatives match across $\rin$; the
Darmois--Israel junction conditions are satisfied and no distributional thin
shell is introduced \cite{Israel1966}.  The matching is verified explicitly
in Sec.~\ref{sec:metric}.

%%%%%%%%%%%%%%%
%%%%%%%%%%%%%%%
%%%%%%%%%%%%%%%

\section{Dressed horizon from absorbed Dehnen mass}
\label{sec:horizon}

This section computes the mass removed from the initial Dehnen seed and uses
the maximal-accretion prescription to determine the dressed horizon.

For $0\leq\gamma<3$, take the initial seed density to be the ordinary
Dehnen profile \cite{Dehnen1993}
\begin{equation}
\rho_\gamma^{(0)}(r)
=\frac{(3-\gamma)\Mh a}
{4\pi r^\gamma(r+a)^{4-\gamma}}.
\label{eq:initial_dehnen_density}
\end{equation}
This is the galaxy-scale input profile before the central depletion is imposed.
It must be distinguished from the final density supporting the static metric.
Integrating Eq.~\eqref{eq:initial_dehnen_density} gives the standard Dehnen
enclosed mass \cite{Dehnen1993}
\begin{equation}
\mathcal M_\gamma^{(0)}(r)
=\Mh\left(\frac{r}{r+a}\right)^{3-\gamma}.
\label{eq:initial_mass}
\end{equation}
In accordance with the maximal-accretion assumption,
%$\varepsilon_{\rm acc}=1$
we assume that the part originally lying inside $\rin$ has been incorporated into the BH:
\begin{equation}
	M_{\rm abs,\gamma}
	=\mathcal M_\gamma^{(0)}(\rin)
	=\Mh u_{\mathrm{in},\gamma}^{3-\gamma},
	\quad
	u_{\mathrm{in},\gamma}\equiv\frac{\rin}{\rin+a}.
	\label{eq:absorbed_mass}
\end{equation}
The dressed mass and horizon are
\begin{equation}
\Meff=\Mbare+\Mh u_{\mathrm{in},\gamma}^{3-\gamma},
\quad
\rh=2\Meff.
\label{eq:dressed_mass_horizon}
\end{equation}
The relation $\rin=\kappa\rh$ and
Eq.~\eqref{eq:dressed_mass_horizon} form the implicit
horizon equation
\begin{equation}
		\rh=2\left[
		\Mbare+\Mh
		\left(\frac{\kappa\rh}{a+\kappa\rh}\right)^{3-\gamma}
		\right].
	\label{eq:horizon_master}
\end{equation}
%The physical branch is the positive solution continuously connected to $\rh=2\Mbare$ as $\Mh\to0$.  Thus the vacuum limit is $\Mh\to0\Rightarrow\rh\to2\Mbare$; a halo parameter cannot alter the Schwarzschild solution after the halo has been removed.

The physically relevant solution is the positive root that continuously approaches $\rh=2\Mbare$ as $\Mh\to0$. Thus, once the halo is removed, the horizon reduces exactly to its Schwarzschild value and retains no dependence on the halo parameters.
For a weak halo, a single expansion of Eq.~\eqref{eq:horizon_master} gives
\begin{equation}
	\rh=2\left[\Mbare+\Mh
	\left(\frac{2\kappa\Mbare}{a+2\kappa\Mbare}\right)^{3-\gamma}\right]
	+\mathcal O(\Mh^2).
	\label{eq:weak_horizon_compact}
\end{equation}
The leading corrections for $\gamma=0,1,2,$ and $5/2$ therefore scale as the
third, second, first, and one-half powers of the quantity in parentheses.
At fixed initial parameters, a steeper cusp places more of the seed halo
inside the prescribed edge and produces a larger leading correction to the
dressed horizon.

\begin{figure}[t]
	\centering
	\includegraphics[width=0.48\textwidth]{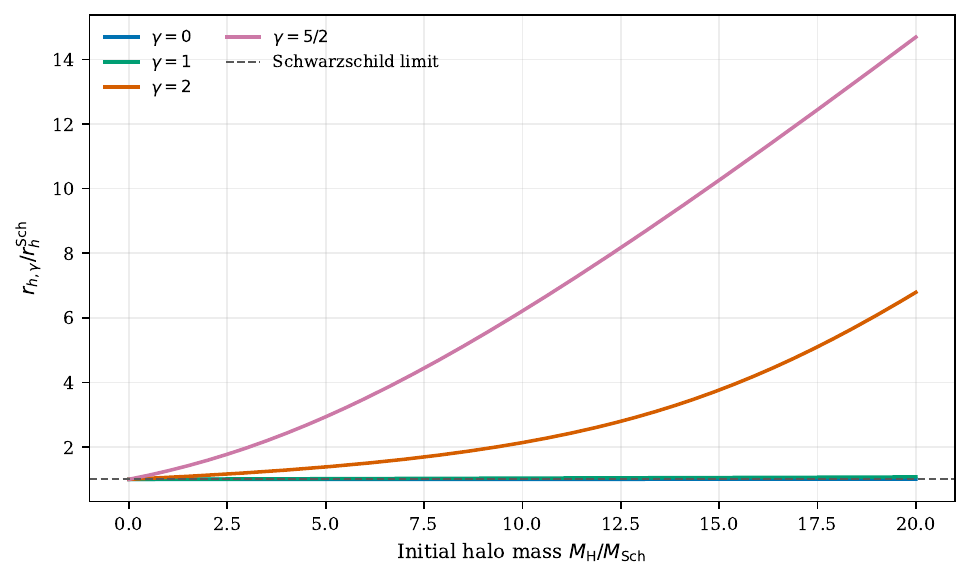}
	\caption{Dressed event-horizon radius relative to the reference
		Schwarzschild value $\rhSch=2\Mbare$, obtained from the implicit mass
		equation~\eqref{eq:horizon_master} for $a/\Mbare=100$, $\kappa=3$, and
		varying initial halo mass.  The plotted root is the branch continuously
		connected to $r_h=2\Mbare$ at $\Mh=0$.  The dashed line is the Schwarzschild
		value in the absence of absorbed halo mass.  At every plotted point the halo
		edge satisfies $r_{\mathrm{in},\gamma}=3r_{h,\gamma}>r_{h,\gamma}$.
		A numerical evaluation of the complete exterior solution over the displayed
		range gives $B_\gamma(r)\geq2/3$ for $r\geq r_{\mathrm{in},\gamma}$, so no
		additional exterior horizon occurs for these parameters.}
	\label{fig:dressed_horizon}
\end{figure}

Figure~\ref{fig:dressed_horizon} displays the physical branch of
Eq.~\eqref{eq:horizon_master} for a representative stable-edge choice
$\kappa=3$.  All curves meet the Schwarzschild value at $\Mh=0$, as required
by the vacuum limit.  At fixed $(\Mbare,\Mh,a,\kappa)$, increasing
$\gamma$ places a larger fraction of the initial Dehnen mass inside $\rin$ and
therefore produces a larger dressed horizon.  The weak response of the cored
and Hernquist profiles in this example is physical rather than numerical: for
$a\gg\rin$, their absorbed fractions scale as
$(\rin/a)^3$ and $(\rin/a)^2$, respectively.

%%%%%%%%%%%%%%%
%%%%%%%%%%%%%%%
%%%%%%%%%%%%%%%

\section{Depleted exterior density and its exact mass}
\label{sec:density}

Having fixed the absorbed mass, we construct and normalize the surviving halo,
then integrate its density to obtain the exterior and total mass functions.

\begin{figure*}[t]
	\centering
	\includegraphics[width=0.92\textwidth]{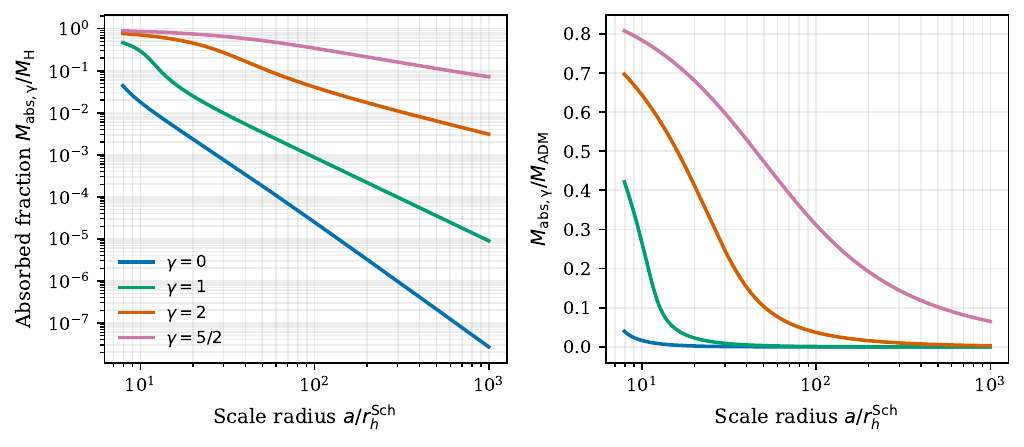}
	\caption{Mass bookkeeping for $\Mh/\Mbare=10$ and $\kappa=3$ as the
		Schwarzschild-normalized scale $a/\rhSch$ is varied.  Left: fraction of the initial halo assigned
		to the dressed BH.  Right: the same absorbed mass relative to the
		conserved ADM mass $M_{\rm ADM}=\Mbare+\Mh$.  Steeper cusps retain a larger
		central mass fraction at a fixed scale radius.}
	\label{fig:mass_bookkeeping}
\end{figure*}

The remaining exterior mass is
\begin{equation}
\Mext=\Mh-M_{\rm abs,\gamma}
=\Mh(1-u_{\mathrm{in},\gamma}^{3-\gamma}).
\label{eq:remaining_mass}
\end{equation}
The division of the conserved initial halo mass into absorbed and exterior
parts is illustrated in Fig.~\ref{fig:mass_bookkeeping}.  The two panels make
clear that changing $\gamma$ does not create mass: it changes only how the
fixed ADM budget $\Mbare+\Mh$ is partitioned between the dressed BH
and the remaining exterior halo.  In the diffuse limit $a\to\infty$, the
absorbed fraction tends to zero for every $\gamma<3$.

Removing the initial mass below $\rin$ does not by itself determine how the
remaining distribution relaxes.  To make this modeling freedom explicit,
consider the normalized family \cite{Shen2025}
\begin{equation}
\rho_{\gamma,n}(r)=
\Theta(r-\rin)
\frac{(3-\gamma)\Mext a}{4\pi J_{\gamma,n}}
\frac{(1-\rin/r)^n}
{r^\gamma(r+a)^{4-\gamma}},
\label{eq:density_n_general}
\end{equation}
where we take $n\geq0$ and
\begin{equation}
J_{\gamma,n}=(3-\gamma)\int_{u_{\mathrm{in},\gamma}}^1
u^{2-\gamma}
\left[
\frac{u-u_{\mathrm{in},\gamma}}{u(1-u_{\mathrm{in},\gamma})}
\right]^n\dd u.
\label{eq:Jpn}
\end{equation}
This normalization guarantees
\begin{equation}
4\pi\int_{\rin}^{\infty}r^2\rho_{\gamma,n}(r)\dd r=\Mext.
\label{eq:density_normalization}
\end{equation}
Two choices have different physical meanings.  For $n=0$,
\begin{equation}
J_{\gamma,0}=1-u_{\mathrm{in},\gamma}^{3-\gamma},
\qquad
\rho_{\gamma,0}(r)=
\Theta(r-\rin)\rho_\gamma^{(0)}(r),
\label{eq:n0_truncated_dehnen}
\end{equation}
so the original galaxy-derived Dehnen density is left unchanged outside a
sharp cutoff.  It has a finite density jump at $\rin$.  For $n=1$, the density
vanishes continuously at the inner edge and reproduces the depletion prescription
used in Ref.~\cite{Shen2025}.  The latter is the analytic model studied in the
rest of this work:
\begin{equation}
		\rho_\gamma(r)=
		\Theta(r-\rin)
		\frac{(3-\gamma)\Mext a}{4\pi J_\gamma}
		\frac{1-\rin/r}{r^\gamma(r+a)^{4-\gamma}},
	\label{eq:density_general}
\end{equation}
for notational simplicity, we henceforth set $J_\gamma\equiv J_{\gamma,1}$ and $\rho_\gamma\equiv \rho_{\gamma,1}$.  
The profile preserves the characteristic Dehnen radial factor and its $r^{-4}$ asymptotic falloff, but it is not simply the original Dehnen density truncated at $r=\rin$. In fact,
\begin{equation}
		\frac{\rho_\gamma(r)}{\rho_\gamma^{(0)}(r)}
		=\frac{1-u_{\mathrm{in},\gamma}^{3-\gamma}}{J_\gamma}
		\left(1-\frac{\rin}{r}\right),
		\qquad r>\rin.
	\label{eq:density_ratio}
\end{equation}
Thus, the $n=1$ prescription entails a redistribution of the surviving
exterior halo in addition to the removal of the original inner mass.
We adopt it because it yields a continuous matter profile, permits smooth matching at the halo edge without a thin shell, and, in the $\gamma=1$ sector and the appropriate parameter limits, recovers the geometries of Refs.~\cite{Shen2025,Cardoso2022}. These properties motivate the choice as an analytic model; they do not constitute a derivation of the relaxed density profile from collisionless dynamics.

For the integrations below, we use the dimensionless radius
$u\equiv r/(r+a)$; its value at the halo edge is
$u_{\mathrm{in},\gamma}$, already defined in
Eq.~\eqref{eq:absorbed_mass}.  The depletion factor is
\begin{equation}
	1-\frac{\rin}{r}
	=\frac{u-u_{\mathrm{in},\gamma}}{u(1-u_{\mathrm{in},\gamma})}.
	\label{eq:depletion_u_relation}
\end{equation}
It is convenient to collect the integrated mass into the single
$\gamma$-dependent function
\begin{widetext}
\begin{equation}
	I_\gamma(u)=\frac{1}{1-u_{\mathrm{in},\gamma}}
	\begin{cases}
		\displaystyle
		u^{3-\gamma}-u_{\mathrm{in},\gamma}^{3-\gamma}
		-\dfrac{(3-\gamma)u_{\mathrm{in},\gamma}}{2-\gamma}
		\left(u^{2-\gamma}-u_{\mathrm{in},\gamma}^{2-\gamma}\right),
		&\gamma\neq2,\\[3mm]
		\displaystyle
		(u-u_{\mathrm{in},\gamma})-u_{\mathrm{in},\gamma}\ln(u/u_{\mathrm{in},\gamma}),
		&\gamma=2.
	\end{cases}
	\label{eq:Igamma}
\end{equation}
\end{widetext}
For each Dehnen model its normalization is defined directly by
\begin{widetext}
\begin{equation}
	J_\gamma\equiv I_\gamma(1)
	=\frac{1}{1-u_{\mathrm{in},\gamma}}
	\begin{cases}
		\displaystyle
		1-u_{\mathrm{in},\gamma}^{3-\gamma}
		-\dfrac{(3-\gamma)u_{\mathrm{in},\gamma}}{2-\gamma}
		\left(1-u_{\mathrm{in},\gamma}^{2-\gamma}\right),
		&\gamma\neq2,\\[3mm]
		\displaystyle
		1-u_{\mathrm{in},\gamma}+u_{\mathrm{in},\gamma}\ln u_{\mathrm{in},\gamma},
		&\gamma=2.
	\end{cases}
	\label{eq:Jgamma}
\end{equation}
\end{widetext}
Direct integration of $h_\gamma'(r)=4\pi r^2\rho_\gamma(r)$ yields
\begin{equation}
h_\gamma(r)=
		\begin{cases}
			0, & r\leq\rin,\\[1mm]
			\displaystyle
			\Mext\frac{I_\gamma(u)}{J_\gamma}, & r>\rin.
	\end{cases}
	\label{eq:h_general}
\end{equation}
Thus,
\begin{equation}
	h_\gamma(\rin)=h_\gamma'(\rin)=0,
	\qquad
	h_\gamma(\infty)=\Mext.
	\label{eq:h_checks}
\end{equation}
No Dirac delta layer is produced at the halo edge.  The total mass function is
\begin{equation}
m_\gamma(r)=\Meff+h_\gamma(r).
	\label{eq:total_mass}
\end{equation}
Its asymptotic value is
\begin{equation}
	m_\gamma(\infty)
	=\Mbare+M_{\rm abs,\gamma}+\Mext
	=\Mbare+\Mh,
	\label{eq:ADMmass}
\end{equation}
so absorption only redistributes the initial total mass and does not change the
ADM mass.  This intrinsic convergence distinguishes the Dehnen construction
from power-law Einstein-cluster toy profiles whose mass functions grow without
bound if extrapolated to infinity.  
%In Ref.~\cite{Maeda2025}, such toy profiles are consistently terminated at a finite outer radius and matched to a constant-mass vacuum exterior, whereas that work's more realistic Model III approaches a finite asymptotic mass without an outer cutoff.  
In Ref.~\cite{Maeda2025}, the power-law models with nonconvergent mass profiles are terminated at a finite outer radius and matched to a constant-mass vacuum exterior, whereas Model~III approaches a finite asymptotic mass without requiring an outer cutoff.
Our profile belongs to the latter finite-mass class because its $r^{-4}$ tail makes the normalization integral in Eq.~\eqref{eq:density_normalization} convergent.

With the Einstein-cluster condition $P_{r,\gamma}=0$ specified, the mass function
$m_\gamma(r)$ provides the complete matter input needed to determine the
metric functions in the next section.

%%%%%%%%%%%%%%%
%%%%%%%%%%%%%%%
%%%%%%%%%%%%%%%

\section{General metric solution}
\label{sec:metric}

We solve the radial Einstein equations once for arbitrary $\gamma$.  The result
provides a universal redshift phase and radial lapse that will be evaluated for
the four representative Dehnen profiles in Sec.~\ref{sec:dehnen-cases}.
Consider the standard static, spherically symmetric areal-radius form
\cite{Poisson2004}
\begin{equation}
	\dd s^2=-A_\gamma(r)\dd t^2
	+\frac{\dd r^2}{B_\gamma(r)}+r^2\dd\Omega^2,
	\label{eq:metric}
\end{equation}
writing $B_\gamma(r)=1-2m_\gamma(r)/r$ identifies $m_\gamma(r)$ as the
Misner--Sharp mass \cite{MisnerSharp1964}.  For an Einstein cluster, the
independent radial equations are
$m_\gamma'(r)=4\pi r^2\rho_\gamma(r)$ and
$A_\gamma'(r)/A_\gamma(r)=
2m_\gamma(r)/\{r[r-2m_\gamma(r)]\}$, with
$P_{r,\gamma}(r)=0$ \cite{Einstein1939,Cardoso2022,Figueiredo2023}.  Using
$m_\gamma(r)=\rh/2+h_\gamma(r)$, the radial metric
function is
\begin{equation}
B_\gamma(r)=1-\frac{2m_\gamma(r)}r
		=\begin{cases}
			1-\dfrac{\rh}{r}, & r\leq\rin,\\[2mm]
			1-\dfrac{\rh+2h_\gamma(r)}{r}, & r>\rin.
	\end{cases}
	\label{eq:B_general}
\end{equation}
The temporal metric function can be written as
\begin{equation}
	A_\gamma(r)=
	\left(1-\frac{\rh}{r}\right)\e^{\Gamma_\gamma(r)},
	\label{eq:Aphase}
\end{equation}
where the exterior phase is determined by imposing asymptotic flatness,
\begin{equation}
\Gamma_\gamma(r)\!=\!
-\! \int_{r}^\infty
\frac{2h_\gamma(x)\,\dd x}
{(x-\rh)[x-\rh-2h_\gamma(x)]},
\quad r\geq\rin.
\label{eq:Gamma_general}
\end{equation}
%In the vacuum gap $\rh<r\leq\rin$, $h_\gamma(r)=0$ and the phase is constant:
Within the vacuum gap, $\rh<r\leq\rin$, one has
$h_\gamma(r)=0$, so the phase reduces to a constant:
\begin{equation}
	\Gamma_\gamma(r)=\Gamma_\gamma(\rin),
	\qquad
	A_\gamma(r)=
	\e^{\Gamma_\gamma(\rin)}
	\left(1-\frac{\rh}{r}\right).
	\label{eq:A_vacuum_gap}
\end{equation}
%Thus $A_\gamma(r)$ and $B_\gamma(r)$ possess the common dressed horizon $\rh$. At the outer edge of the gap, a complete junction requires the matching of both metric functions, not only the redshift function.  In the present coordinates the Darmois--Israel conditions \cite{Israel1966} reduce to
%\begin{equation}
%	\begin{aligned}
%		A_{\rm gap}(\rin)&=A_{\rm ext}(\rin),&
%		B_{\rm gap}(\rin)&=B_{\rm ext}(\rin),\\
%		A_{\rm gap}'(\rin)&=A_{\rm ext}'(\rin),&
%		B_{\rm gap}'(\rin)&=B_{\rm ext}'(\rin).
%	\end{aligned}
%	\label{eq:full_edge_matching}
%\end{equation}
%The first equality fixes the constant Killing-time normalization $C_\gamma=\exp[\Gamma_\gamma(\rin)]$.  The remaining equalities follow from $h_\gamma(\rin)=h_\gamma'(\rin)=0$ for the $n=1$ profile.  Hence both the temporal redshift function $A_\gamma(r)$ and the radial lapse $B_\gamma(r)$ join smoothly, and no thin shell is hidden at $r=\rin$.
Thus both metric functions $A_\gamma(r)$ and $B_\gamma(r)$ vanish at $r=\rh$, identifying the common dressed horizon. At the halo edge $r=\rin$, the absence of a distributional surface layer requires the induced metric and the extrinsic curvature to be continuous across the junction. In the present areal-radius coordinates, the Darmois--Israel conditions \cite{Israel1966} become
\begin{eqnarray}	\label{eq:full_edge_matching}
		A_{\rm gap}(\rin)&=&A_{\rm ext}(\rin),\\
		B_{\rm gap}(\rin)&=&B_{\rm ext}(\rin),\\
		A_{\rm gap}'(\rin)&=&A_{\rm ext}'(\rin).
	\end{eqnarray}
Continuity of $A_\gamma$ determines the constant Killing-time
normalization $C_\gamma=\exp[\Gamma_\gamma(\rin)]$.
For the $n=1$ profile, $h_\gamma(\rin)=h_\gamma'(\rin)=0$ ensures the continuity of $B_\gamma$ and, in addition, gives
\begin{equation}
	B_{\rm gap}'(\rin)=B_{\rm ext}'(\rin).
\end{equation}
Hence the metric functions join smoothly at the halo edge, and no thin shell is generated at $r=\rin$.

%%%%%%%%%%%%%%%
%%%%%%%%%%%%%%%
%%%%%%%%%%%%%%%

\section{Explicit Dehnen cases}
\label{sec:dehnen-cases}

We now specialize the general mass function, radial lapse, and redshift phase
to $\gamma=0,1,2,$ and $5/2$.  The following subsections proceed from the
cored model to the steep cusp.  For each case we first obtain
$h_\gamma(r)$ and $B_\gamma(r)$ and then determine the exterior redshift
function $A_\gamma(r)$; the common vacuum-gap expression remains
Eq.~\eqref{eq:A_vacuum_gap} and is not repeated.

\subsection{Cored profile: \texorpdfstring{$\gamma=0$}{gamma=0}}

For $\gamma=0$, the normalization factor reduces to $J_0=(1-u_{\mathrm{in},\gamma})(2+u_{\mathrm{in},\gamma})/2$.
Using the relation between $u$ and $r$ in
Eq.~\eqref{eq:depletion_u_relation}, the general mass function
\eqref{eq:h_general} takes the areal-radius form
\begin{equation}
		h_0(r)=\Mext
		\frac{(r-\rin)^2[(3\rin+2a)r+a\rin]}
		{(r+a)^3(3\rin+2a)}.
	\label{eq:h0r}
\end{equation}
It is convenient to introduce
\begin{eqnarray}
	Q_0(r)&=&(3\rin+2a)(r+a)^3,
	\label{eq:Q0}\\
	\nonumber
	N_0(r)&=&\Mext(r-\rin)^2\\ &&[(3\rin+2a)r+a\rin],
	\label{eq:N0}\\
	P_0(r)&=&Q_0(r)(r-\rh)-2N_0(r).
	\label{eq:P0}
\end{eqnarray}
The exact radial function is
\begin{equation}
	B_0(r)=\frac{P_0(r)}{rQ_0(r)}\quad (r\geq\rin).
	\label{eq:B0}
\end{equation}
Assuming that the four roots $z_i^{(0)}$ of $P_0(r)$ are simple, define
\begin{equation}
	c_i^{(0)}=\frac{Q_0(z_i^{(0)})}{P_0'(z_i^{(0)})}.
	\label{eq:c0}
\end{equation}
The radial Einstein equation can then be written as
\begin{equation}
	\frac{A_0'}{A_0}
	=-\frac1r+\frac{Q_0(r)}{P_0(r)},
\end{equation}
Integrating this expression and imposing asymptotic normalization yields the exact exterior temporal metric function
\begin{equation}
		A_0(r)=\prod_{i=1}^{4}
		\left(1-\frac{z_i^{(0)}}r\right)^{c_i^{(0)}},
		\qquad r\geq\rin.
	\label{eq:A0roots}
\end{equation}
The residues satisfy $\sum_i c_i^{(0)}=1$. Any nonreal roots and the corresponding exponents occur in complex-conjugate pairs, so their contributions combine to give a real metric function on the physical exterior domain.

\subsection{Hernquist profile: \texorpdfstring{$\gamma=1$}{gamma=1}}

Here the normalization factor becomes $J_1=1-u_{\mathrm{in},\gamma}$.  Using again
Eq.~\eqref{eq:depletion_u_relation}, the general mass reduces directly to
\begin{equation}
		h_1(r)=\Mext\left(\frac{r-\rin}{r+a}\right)^2.
	\label{eq:h1}
\end{equation}
This is precisely the shifted Hernquist mass appearing in the special model of Ref.~\cite{Shen2025}.  It is convenient to define
\begin{eqnarray}
	Q_1(r)&=&(r+a)^2,\\
	P_1(r)&=&(r+a)^2(r-\rh)\\&&-2\Mext(r-\rin)^2.
	\label{eq:P1}
\end{eqnarray}
The radial metric function then takes the form
\begin{equation}
B_1(r)=\frac{P_1(r)}{r(r+a)^2},\qquad r\geq\rin.
	\label{eq:B1}
\end{equation}
If $z_i^{(1)}$, $i=1,2,3$, are the simple roots of $P_1$, set
\begin{equation}
	c_i^{(1)}=
	\frac{\bigl(z_i^{(1)}+a\bigr)^2}{P_1'(z_i^{(1)})}.
	\label{eq:c1}
\end{equation}
The exact exterior redshift function is
\begin{equation}
		A_1(r)=\prod_{i=1}^{3}
		\left(1-\frac{z_i^{(1)}}r\right)^{c_i^{(1)}},
		\qquad r\geq\rin.
	\label{eq:A1roots}
\end{equation}
Here $\sum_i c_i^{(1)}=1$, and complex-conjugate factors combine to give a real function.  This representation also makes asymptotic flatness transparent.  Since $h_1(r)=\Mext+O(r^{-1})$, the total mass approaches $M_{\rm ADM}=\Meff+\Mext=\Mbare+\Mh$, and therefore
\begin{equation}
	B_1(r)=1-\frac{2M_{\rm ADM}}{r}+O(r^{-2}).
	\label{eq:B1_asymptotic}
\end{equation}
For the redshift function,
\begin{align}
	A_1(r)
	&=\prod_{i=1}^{3}\left(1-\frac{z_i^{(1)}}r\right)^{c_i^{(1)}} \notag\\
	&=1-\frac{\sum_i c_i^{(1)}z_i^{(1)}}r+O(r^{-2})\notag\\
	&=1-\frac{2M_{\rm ADM}}r+O(r^{-2}),
	\label{eq:A1_asymptotic}
\end{align}
where the last identity follows equivalently by expanding
$A_1'/A_1=2m_1/[r(r-2m_1)]$ at infinity.  Thus
$A_1\to1$ and $B_1\to1$, while the angular sector already has the standard
areal-radius form.  The $\gamma=1$ geometry is therefore asymptotically
Minkowskian; the same conclusion follows from the general phase integral for
every finite-mass member of the family.

The same integrated $\gamma=1$ solution contains the geometries of
Refs.~\cite{Shen2025,Cardoso2022}; no independent radial integration is required.  To reproduce the notation of Ref.~\cite{Shen2025}, let $M_{\rm BH}$ and $M_h$ be the final BH and surviving-halo masses, set $\nu_{\rm SWY}=2\kappa$, and define $u_{\rm SWY}=\nu_{\rm SWY}M_{\rm BH}/(a+\nu_{\rm SWY}M_{\rm BH})$.
Their fixed-final-mass parametrization is related to our initial data by
\begin{equation}
	\Mh=\frac{M_h}{1-u_{\rm SWY}^{2}},\qquad
	\Mbare=M_{\rm BH}-\frac{M_hu_{\rm SWY}^{2}}{1-u_{\rm SWY}^{2}}.
	\label{eq:plb_input_map_compact}
\end{equation}
Substitution in Eqs.~\eqref{eq:density_general} and \eqref{eq:h1} immediately
gives
\begin{widetext}
\begin{eqnarray}
	\rho_1(r)&=&\Theta(r-\nu_{\rm SWY}M_{\rm BH})
	\frac{M_h(a+\nu_{\rm SWY}M_{\rm BH})}{2\pi r(r+a)^3}
	\left(1-\frac{\nu_{\rm SWY}M_{\rm BH}}r\right),\nonumber\\
	m_1(r)&=&M_{\rm BH}+\Theta(r-\nu_{\rm SWY}M_{\rm BH})M_h
	\frac{(r-\nu_{\rm SWY}M_{\rm BH})^2}{(r+a)^2},
	\label{eq:plb_source_geometry_compact}
\end{eqnarray}
\end{widetext}
which are the source and mass function of Ref.~\cite{Shen2025}.  Specializing the already-integrated lapse \eqref{eq:A1roots} gives their closed-form redshift function; in the astrophysical hierarchy the associated cubic has one real root and a complex-conjugate pair whose contributions combine into a real lapse.  
The apparent difference in the horizon prescriptions therefore arises from the distinct mass parametrizations: Ref.~\cite{Shen2025} takes the final BH and halo masses as fixed inputs, whereas our construction is specified by $(\Mbare,\Mh)$ prior to depletion.

The no-gap endpoint $\kappa=1$, with $\rin=\rh=2M_{\rm BH}$,
$\Mext=M$, and $a=a_0$, yields the geometry of Ref.~\cite{Cardoso2022}:
\begin{align}
	m_{\rm C}(r)&=M_{\rm BH}+M\frac{(r-2M_{\rm BH})^2}{(r+a_0)^2},\nonumber\\
	B_{\rm C}(r)&=\left(1-\frac{2M_{\rm BH}}r\right)
	\left[1-\frac{2M(r-2M_{\rm BH})}{(r+a_0)^2}\right],\nonumber\\
	A_{\rm C}(r)&=\left(1-\frac{2M_{\rm BH}}r\right)e^{\Gamma_{\rm C}(r)},
	\label{eq:cardoso_mass_B_compact}
\end{align}
where, with $\xi=2a_0-M+4M_{\rm BH}$,
\begin{equation}
	\Gamma_{\rm C}(r)=-\pi\sqrt{\frac M\xi}
	+2\sqrt{\frac M\xi}\arctan \left(
	\frac{r+a_0-M}{\sqrt{M\xi}}\right).
	\label{eq:cardoso_phase_compact}
\end{equation}
The corresponding density is the $\gamma=1$, $\rin=2M_{\rm BH}$ limit of
Eq.~\eqref{eq:density_general} and vanishes at the horizon.  Nevertheless,
$v_{\rm C}^{2}=m_{\rm C}/(r-2m_{\rm C})\sim
M_{\rm BH}/(r-2M_{\rm BH})$ diverges as the horizon is approached.  Hence the
no-gap source of Ref.~\cite{Cardoso2022} is a valid effective anisotropic
fluid, but its near-horizon
part cannot literally consist of massive particles on circular timelike
orbits.  This physical distinction motivates the finite vacuum gap in the
present construction.

\subsection{Jaffe profile: \texorpdfstring{$\gamma=2$}{gamma=2}}

Here
$J_2=(1-u_{\mathrm{in},\gamma}+u_{\mathrm{in},\gamma}\ln u_{\mathrm{in},\gamma})/(1-u_{\mathrm{in},\gamma})$, and the
normalized enclosed exterior mass is logarithmic:
\begin{equation}
		h_2(u)=\Mext
		\frac{(u-u_{\mathrm{in},\gamma})-u_{\mathrm{in},\gamma}\ln(u/u_{\mathrm{in},\gamma})}
		{1-u_{\mathrm{in},\gamma}+u_{\mathrm{in},\gamma}\ln u_{\mathrm{in},\gamma}}.
	\label{eq:h2}
\end{equation}
It is convenient to introduce
\begin{align}
	\ell(u)&=(a+\rh)u-\rh,\\
	\mathcal D_2(u)&=
	\ell(u)-2(1-u)h_2(u).
	\label{eq:D2}
\end{align}
The exterior metric functions then take the exact form
\begin{equation}
	\resizebox{0.85\columnwidth}{!}{$
		\begin{aligned}
			B_2(u)
			&=\frac{\mathcal D_2(u)}{au},\\
			A_2(r)
			&=\left(1-\frac{\rh}{r}\right)
			\exp\left[
			-\int_{u(r)}^1
			\frac{2a\,h_2(v)\,\dd v}
			{\ell(v)\mathcal D_2(v)}
			\right],
		\end{aligned}
		$}
	\label{eq:A2}
\end{equation}
with $r\geq\rin$.
Because $h_2(u)$ contains a logarithmic term, the integral entering
$A_2(r)$ does not generally admit a useful closed-form expression in
elementary functions. Equation~\eqref{eq:A2} nevertheless provides an exact one-dimensional quadrature and is particularly convenient for numerical evaluation, avoiding the need for a more cumbersome
special-function representation.

\subsection{Steep cusp: \texorpdfstring{$\gamma=5/2$}{gamma=5/2}}

For this case $J_{5/2}=(1-y_{\rm in})/(1+y_{\rm in})$.  Considering
\begin{equation}
	y=\sqrt{u}=\sqrt{\frac{r}{r+a}},
	\qquad
	y_{\rm in}=\sqrt{u_{\mathrm{in},\gamma}}.
	\label{eq:ydef}
\end{equation}
the exact exterior mass can be written as
\begin{equation}
		h_{5/2}(y)=\Mext
		\frac{(y-y_{\rm in})^2}
		{y(1-y_{\rm in})^2}.
	\label{eq:h52}
\end{equation}
We introduce
\begin{eqnarray}
	L(y)&=&(a+\rh)y^2-\rh,
	\label{eq:L52}\\ \nonumber
	P_{5/2}(y)&=&
	y(1-y_{\rm in})^2L(y)\\ 
	&&-2\Mext(1-y^2)(y-y_{\rm in})^2.
	\label{eq:P52}
\end{eqnarray}
Then
\begin{equation}
		B_{5/2}(y)=
		\frac{P_{5/2}(y)}{a y^3(1-y_{\rm in})^2},
		\qquad y\geq y_{\rm in}.
	\label{eq:B52}
\end{equation}
The phase is the rational integral
\begin{equation}
		\Gamma_{5/2}(y)=
		-\int_y^1
		\frac{4a\Mext z(z-y_{\rm in})^2\,\dd z}
		{L(z)P_{5/2}(z)}.
	\label{eq:Gamma52}
\end{equation}
Let $\zeta_j$ denote the six generic simple roots of
$L(y)P_{5/2}(y)$, and define the residues
\begin{equation}
	d_j=
	\frac{4a\Mext\zeta_j(\zeta_j-y_{\rm in})^2}
	{\left.\dfrac{\dd}{\dd y}
		[L(y)P_{5/2}(y)]\right|_{y=\zeta_j}}.
	\label{eq:d52}
\end{equation}
Partial fractions give
\begin{equation}
		\Gamma_{5/2}(y)=
		\sum_{j=1}^{6}d_j
		\ln \left(\frac{y-\zeta_j}{1-\zeta_j}\right),
	\label{eq:Gamma52roots}
\end{equation}
and therefore
\begin{equation}
	A_{5/2}(r)=
		\left(1-\frac{\rh}{r}\right)
		\prod_{j=1}^{6}
		\left(\frac{\sqrt{r/(r+a)}-\zeta_j}{1-\zeta_j}\right)^{d_j},
	\label{eq:A52}
\end{equation}
by considering $r\geq\rin$.
Consistent logarithm branches make the contributions from complex-conjugate
roots combine to a real lapse.

The explicit solutions must finally satisfy the common horizon test.  Because
$\kappa>1$, the dressed horizon lies inside the vacuum gap.  In
$\rh<r\leq\rin$, one has $B_\gamma(r)=1-\rh/r>0$; hence no additional
horizon occurs in this region.  In the exterior halo, the absence of an
additional zero of $B_\gamma(r)$ is equivalent to the exact condition
\begin{equation}
		r-\rh-2h_\gamma(r)>0,
		\qquad r\geq\rin.
	\label{eq:noextra_general}
\end{equation}
For the explicit profiles, this is the positivity of $P_0(r)$ for $\gamma=0$,
$P_1(r)$ for $\gamma=1$, $\mathcal D_2(u)$ for $\gamma=2$, and
$P_{5/2}(y)$ for $\gamma=5/2$ over their respective exterior domains.  This
exact condition, rather than zeros of a truncated expansion of $A_\gamma(r)$,
must be tested for every parameter set.  When it holds, the denominators in
the phase integrals remain finite and $\exp[\Gamma_\gamma(r)]>0$.  Therefore,
the only common zero of the two metric functions is
$A_\gamma(\rh)=B_\gamma(\rh)=0$; no further horizon occurs for $r>\rh$.

%%%%%%%%%%%%%%%
%%%%%%%%%%%%%%%
%%%%%%%%%%%%%%%

\section{Physical admissibility: energy conditions and orbital stability}
\label{sec:admissibility}

We next test whether the exterior source is physically admissible.  We first
derive the energy-condition bounds for the effective anisotropic fluid and
then impose the stronger timelike-orbit and stability requirements needed for
a literal Einstein-cluster interpretation.

\subsection{Energy conditions}

The exterior matter source is modeled as an Einstein cluster
\cite{Einstein1939,Cardoso2022,Acharyya2024},
\begin{equation}
	T^{\mu}{}_{\nu}
	=
	\operatorname{diag}
	\left(
	-\rho_\gamma(r),\,0,\,
	P_{t,\gamma}(r),\,P_{t,\gamma}(r)
	\right),
	\label{eq:stress_tensor}
\end{equation}
with vanishing radial pressure, $P_{r,\gamma}=0$.
Stress-energy conservation, equivalently the angular Einstein equation,
gives the standard Einstein-cluster tangential pressure
\cite{Einstein1939,Cardoso2022},
\begin{equation}
	P_{t,\gamma}(r)
	=
	\frac{m_\gamma(r)\rho_\gamma(r)}
	{2\left[r-2m_\gamma(r)\right]}.
	\label{eq:tangential_pressure}
\end{equation}

In the vacuum gap, both $\rho_\gamma(r)$ and $P_{t,\gamma}(r)$ vanish.
In the halo region, we assume the no-additional-horizon condition
\eqref{eq:noextra_general}, so that $r>2m_\gamma(r)$. Using the standard
pointwise definitions of the energy conditions
\cite{HawkingEllis1973}, the radial null energy condition reduces to
\begin{equation}
	\rho_\gamma(r)+P_{r,\gamma}(r)
	=
	\rho_\gamma(r)\geq0,
\end{equation}
while the tangential null energy condition becomes
\begin{equation}
	\rho_\gamma(r)+P_{t,\gamma}(r)
	=
	\rho_\gamma(r)
	\frac{2r-3m_\gamma(r)}
	{2\left[r-2m_\gamma(r)\right]}
	\geq0.
\end{equation}
The strong-energy combination is
\begin{equation}
	\rho_\gamma(r)+P_{r,\gamma}(r)+2P_{t,\gamma}(r)
	=
	\rho_\gamma(r)
	\frac{r-m_\gamma(r)}
	{r-2m_\gamma(r)}
	\geq0.
	\label{eq:SEC_explicit}
\end{equation}
For $\rho_\gamma(r)\geq0$ and $r>2m_\gamma(r)$, these conditions are
automatically satisfied and therefore impose no additional lower bound
on $\kappa$.

The nontrivial restriction arises from the dominant energy condition
(DEC). Since $P_{r,\gamma}=0$ and $P_{t,\gamma}\geq0$ in the
no-additional-horizon region, the only nontrivial DEC requirement is
$\rho_\gamma(r)\geq P_{t,\gamma}(r)$. At every point where
$\rho_\gamma(r)>0$, this condition is equivalent to
\begin{equation}
	\frac{m_\gamma(r)}
	{2\left[r-2m_\gamma(r)\right]}
	\leq1,
	\label{eq:DEC_general1}
\end{equation}
or equivalently
\begin{equation}
	\mathcal F_\gamma(r;\kappa)
	\equiv
	2r-5m_\gamma(r)\geq0.
	\label{eq:DEC_genera2}
\end{equation}
Moreover, the DEC therefore requires
\begin{equation}
	\frac{m_\gamma(r)}{r}\leq\frac{2}{5}.
\end{equation}
As a direct consequence, the radial metric function satisfies
\begin{equation}
	B_\gamma(r)
	=
	1-\frac{2m_\gamma(r)}{r}
	\geq\frac{1}{5}.
	\label{eq:DEC_implies_no_horizon}
\end{equation}
Hence, whenever the DEC holds throughout the halo, the radial metric
function remains strictly positive there and no additional exterior
horizon can occur.

Thus a globally DEC-admissible halo automatically contains no additional zero
of $B_\gamma(r)$; the remaining independent horizon check is needed only when the
DEC is not imposed or is violated.
Using $m_\gamma(r)=\rh/2+h_\gamma(r)$, this function is
\begin{equation}
	\mathcal F_\gamma(r;\kappa)
	=2r-\frac52\rh-5h_\gamma(r).
	\label{eq:Fgamma}
\end{equation}

Although the density vanishes exactly at $r=\rin$, the DEC must hold in a
one-sided neighborhood immediately above the edge.  Because
$h_\gamma(\rin)=0$, its limiting value is
\begin{equation}
	\mathcal F_\gamma(\rin^+)
	=\left(2\kappa-\frac52\right)\rh.
	\label{eq:DEC_edge}
\end{equation}
Consequently every profile obeys the universal necessary bound
\begin{equation}
\kappa\geq\frac54.
	\label{eq:kappa_edge_bound}
\end{equation}
This edge result is independent of $\gamma$ because no exterior halo mass has
yet accumulated at $r=\rin$.  It must not be confused with the exact global
bound.

For fixed inputs $(\Mbare,\Mh,a)$, define the exact admissible value by
\begin{equation}
		\kappa_{\rm DEC,\gamma}
		=\inf\left\{\kappa>1:\
		\min_{r\geq\kappa\rh(\kappa,\gamma)}
		\mathcal F_\gamma(r;\kappa)\geq0\right\},
	\label{eq:kappa_DEC_definition}
\end{equation}
where $\rh(\kappa,\gamma)$ is the physical solution of
Eq.~\eqref{eq:horizon_master}.  Thus $\kappa_{\rm DEC,\gamma}$ is not a
universal function of $\gamma$ alone; it also depends on the dimensionless
ratios $\Mh/\Mbare$ and $a/\Mbare$.

The derivative of the exact exterior mass is particularly useful.  With
$u=r/(r+a)$, Eqs.~\eqref{eq:density_general} and
\eqref{eq:h_general} give
\begin{equation}
		h_\gamma'(r)
		=\frac{(3-\gamma)\Mext}{aJ_\gamma(1-u_{\mathrm{in},\gamma})}
		(u-u_{\mathrm{in},\gamma})u^{1-\gamma}(1-u)^2.
	\label{eq:hprime_u}
\end{equation}
Since
\begin{equation}
	\mathcal F_\gamma'(r)=2-5h_\gamma'(r),
	\label{eq:Fprime}
\end{equation}
a simple sufficient condition is
\begin{equation}
	\max_{r\geq\rin}h_\gamma'(r)\leq\frac25
	\quad\Longrightarrow\quad
\kappa_{\rm DEC,\gamma}=\frac54.
	\label{eq:monotone_DEC}
\end{equation}
The location $u_*$ of the maximum of $h_\gamma'(r)$ is obtained algebraically
from
\begin{equation}
(4-\gamma)u_*^2-
		\left[2-\gamma+(3-\gamma)u_{\mathrm{in},\gamma}\right]u_*
		+(1-\gamma)u_{\mathrm{in},\gamma}=0,
	\label{eq:u_star}
\end{equation}
where the root in $(u_{\mathrm{in},\gamma},1)$ is selected.  Its coefficients differ for
$\gamma=0,1,2,$ and $5/2$, providing a rapid profile-by-profile test.

If the maximum exceeds $2/5$, the global minimum of
$\mathcal F_\gamma$ may occur at an interior radius $r_c$.  The limiting DEC
configuration is then determined by the coupled system
\begin{equation}
		\begin{aligned}
			\rh&=2\left[\Mbare+\Mh
			\left(\frac{\kappa\rh}{a+\kappa\rh}\right)^{3-\gamma}\right],\\
			2r_c&=5\left[\frac{\rh}{2}+h_\gamma(r_c)\right],\\
			h_\gamma'(r_c)&=\frac25,
			\qquad r_c>\kappa\rh.
	\end{aligned}
	\label{eq:kappa_critical_system}
\end{equation}
Substituting Eqs.~\eqref{eq:h0r}, \eqref{eq:h1}, \eqref{eq:h2}, or
\eqref{eq:h52} into the general derivative~\eqref{eq:hprime_u} gives the four exact
constraint systems for $\kappa_{\rm DEC,0}$,
$\kappa_{\rm DEC,1}$, $\kappa_{\rm DEC,2}$, and
$\kappa_{\rm DEC,5/2}$.  They are generally solved numerically because
$\rh$, $u_{\mathrm{in},\gamma}$, $\Mext$, and $J_\gamma$ also depend on $\kappa$.

In practice, one solves the dressed-horizon equation for a trial $\kappa$ and
then minimizes $\mathcal F_\gamma$ over the full occupied interval.  The
minimization is essential because the DEC is an inequality on a radial domain,
not a second pointwise algebraic equation.  If the minimum occurs at the edge,
the critical value is $5/4$; if it occurs in the matter region, the coupled
system~\eqref{eq:kappa_critical_system} gives a profile-dependent bound.  This
statement assumes that the initial data $(\Mbare,\Mh,a)$ are held fixed, as in
the horizon construction used throughout this work.

The compact choice $a\sim r_h$ can make the interior minimum relevant, but it
is not representative of a galactic halo.  In the intended hierarchy
$a\gg \Mbare,\Mh$, the normalized mass gradient~\eqref{eq:hprime_u} is small
throughout the exterior.  Consequently $\max h_\gamma'(r)\ll2/5$ and
$\mathcal F_\gamma$ increases outward from the halo edge.  The global DEC
condition therefore reduces to the universal edge bound
$\kappa_{\rm DEC,\gamma}=5/4$ for each of the four Dehnen models.

%To verify that this edge condition also guarantees the DEC throughout the occupied halo, we evaluate the global minimum of $\mathcal F_\gamma(r;\kappa)$ for all four profiles.  We use the Schwarzschild-normalized inputs $\Mh/\Mbare=10$ and $a/\Mbare=1000$. Although the total mass budget is halo dominated, the large scale radius keeps most of that mass outside the strong-field region.  For each $\gamma$, the critical $\kappa$ is found by requiring the minimum of $\mathcal F_\gamma$ over the occupied halo to be nonnegative:
%\begin{center}
%	\footnotesize
%	\begin{tabular}{@{}cccccc@{}}
%		\toprule
%		$\gamma$ & $\kappa_\gamma$ & $\rh/\rhSch$ & $\rin/\rhSch$
%		& $M_{\rm abs,\gamma}/\Mh$ & $\Mext/\Mh$\\
%		\midrule
%		$0$   & $1.25$ & $1.000000155$ & $1.250000194$ & $1.551\!\times\!10^{-8}$ & $0.999999985$\\
%		$1$   & $1.25$ & $1.000062197$ & $1.250077746$ & $6.220\!\times\!10^{-6}$ & $0.999993780$\\
%		$2$   & $1.25$ & $1.025573775$ & $1.281967219$ & $2.557\!\times\!10^{-3}$ & $0.997442623$\\
%		$5/2$ & $1.25$ & $1.638763936$ & $2.048454920$ & $6.388\!\times\!10^{-2}$ & $0.936123606$\\
%		\bottomrule
%	\end{tabular}
%\end{center}
%The DEC minimum occurs at the edge, where $B_\gamma(\rin)=0.2$, and no additional horizon is present.  

To determine whether the local edge bound also ensures the DEC throughout
the occupied halo for the diffuse benchmark considered here, we evaluate
the global minimum of $\mathcal F_\gamma(r;\kappa)$ for all four profiles. We take $\Mh/\Mbare=10$ and $a/\Mbare=1000$.
Although the total mass is halo dominated, the large scale radius makes the halo diffuse in the strong-field region. For each $\gamma$, the critical value $\kappa_{\rm DEC,\gamma}$ is obtained by requiring the minimum of $\mathcal F_\gamma(r;\kappa)$ over the full occupied halo to be nonnegative. The resulting values are summarized in
Table~\ref{tab:DEC_global}.
\begin{table}[t]
	\centering
	\caption{
		Global DEC threshold for the four depleted Dehnen profiles with
		$\Mh/\Mbare=10$ and $a/\Mbare=1000$.
		For each $\gamma$, the minimum of
		$\mathcal F_\gamma(r;\kappa)$ over the occupied halo occurs at the
		inner edge, yielding the critical value
		$\kappa_{\rm DEC,\gamma}=5/4$.
		The remaining columns show the corresponding dressed-horizon radius,
		halo-edge radius, absorbed mass fraction, and surviving exterior mass
		fraction.
	}
	\label{tab:DEC_global}
	\footnotesize
	\begin{tabular}{@{}cccccc@{}}
		\toprule
		$\gamma$
		& $\kappa_{\rm DEC,\gamma}$
		& $\rh/\rhSch$
		& $\rin/\rhSch$
		& $M_{\rm abs,\gamma}/\Mh$
		& $\Mext/\Mh$
		\\
		\midrule
		$0$
		& $1.25$
		& $1.000000155$
		& $1.250000194$
		& $1.551\times10^{-8}$
		& $0.999999985$
		\\
		$1$
		& $1.25$
		& $1.000062197$
		& $1.250077746$
		& $6.220\times10^{-6}$
		& $0.999993780$
		\\
		$2$
		& $1.25$
		& $1.025573775$
		& $1.281967219$
		& $2.557\times10^{-3}$
		& $0.997442623$
		\\
		$5/2$
		& $1.25$
		& $1.638763936$
		& $2.048454920$
		& $6.388\times10^{-2}$
		& $0.936123606$
		\\
		\bottomrule
	\end{tabular}
\end{table}
As shown in Table~\ref{tab:DEC_global}, the global minimum of $\mathcal F_\gamma$ occurs at the halo edge for all four profiles, so that $\kappa_{\rm DEC,\gamma}=5/4$. At the critical edge,
$B_\gamma(\rin)=1/5$, and the radial metric function remains positive throughout the exterior region. Hence no additional exterior horizon occurs for these parameter choices.

The maxima of $h_\gamma'(r)$ are
$1.873\times10^{-3}$, $2.956\times10^{-3}$, $8.767\times10^{-3}$, and
$3.141\times10^{-2}$ in increasing order of $\gamma$, all well below $2/5$.
Thus the global calculation confirms $\kappa_{\rm DEC,\gamma}=5/4$ for all
four diffuse profiles, while $\rh$, $\rin$, and the absorbed mass remain
cusp dependent.  Profile-dependent DEC thresholds can nevertheless arise for
compact halos and are then determined by
Eq.~\eqref{eq:kappa_critical_system}.  Having established the admissibility of
the effective continuum source, we next impose the stronger conditions needed
for its interpretation as a cluster of massive particles.

\subsection{Timelike circular motion and cluster stability}

This energy-condition analysis concerns the continuum source.  In the
Einstein-cluster interpretation \cite{Einstein1939}, the constituent particles
move on randomly oriented tangential orbits, with counter-rotation ensuring
that the mean angular momentum and energy flux vanish.  Their quadratic
angular momentum nevertheless produces the nonzero tangential pressure
$P_{t,\gamma}(r)$.  The averaged stress tensor and geometry can consequently
be time independent even though the individual particles are moving.  The
exact solution should therefore be understood as a continuum or mean-field
description; a finite particle realization would exhibit small fluctuations
about it.  This macroscopic equilibrium does not, however, establish the
existence or stability of the assumed circular trajectories.

Energy conditions do not by themselves establish the existence or stability
of circular particle orbits.  For a circular equatorial geodesic, the radial
geodesic equation gives
$\Omega^2=(\dd\phi/\dd t)^2=A_\gamma'(r)/(2r)$.  A static observer measures
$v^2(r)=r^2\Omega^2/A_\gamma(r)$, while the radial Einstein equation with
$P_{r,\gamma}(r)=0$ gives
$A_\gamma'(r)/A_\gamma(r)=
2m_\gamma(r)/\{r[r-2m_\gamma(r)]\}$.  Finally,
$\nabla_\mu T^\mu{}_r=0$ yields
$P_{t,\gamma}(r)=\rho_\gamma(r)rA_\gamma'(r)/[4A_\gamma(r)]$.
Combining these standard relations gives
\begin{eqnarray}
v^2(r)&=&\frac{rA_\gamma'(r)}{2A_\gamma(r)}
		=\frac{m_\gamma(r)}{r-2m_\gamma(r)},\\
P_{t,\gamma}(r)&=&\frac12\rho_\gamma(r)v^2(r).
	\label{eq:orbital_speed}
\end{eqnarray}
The stress-energy trace consequently reduces to
$T^\mu{}_{\mu}=-\rho_\gamma+2P_{t,\gamma}
=\rho_\gamma(v^2-1)$.  It is negative in the timelike-cluster domain,
vanishes at the limiting null orbit, and becomes positive when the formal
circular speed is superluminal.  Thus the trace provides a compact diagnostic
of the same transition, but no independent bound on $\kappa$.
A circular worldline is timelike only if $v^2(r)<1$, equivalently
$r>3m_\gamma(r)$.
At the one-sided halo edge, $m_\gamma(\rin)=\rh/2$, so a literal cluster of
timelike circular particles requires $\kappa>3/2$.
This is stronger than the DEC edge value $5/4$.  Hence the interval
$5/4\leq\kappa\leq3/2$ can describe an effective anisotropic fluid satisfying
the DEC, but not a literal cluster of massive particles on circular timelike
orbits at its inner edge.
Indeed, the exact one-sided edge velocity is
\begin{equation}
	\boxed{v_{\rm edge}^{2}
		=\frac{m_\gamma(\rin)}{\rin-2m_\gamma(\rin)}
		=\frac{1}{2(\kappa-1)}.}
	\label{eq:edge_velocity}
\end{equation}
It takes the values $2$, $1$, and $1/4$ for
$\kappa=5/4$, $3/2$, and $3$, respectively.  The distinct physical roles of
these three values are summarized after imposing the stability condition.

For the general static spherical metric, the energy and angular momentum of a
circular timelike geodesic take the standard form
\cite{Chandrasekhar1983}
\begin{eqnarray}
	E^2&=&\frac{2[A_\gamma(r)]^2}
	{2A_\gamma(r)-rA_\gamma'(r)},\\
	L^2&=&\frac{r^3A_\gamma'(r)}
	{2A_\gamma(r)-rA_\gamma'(r)}
	=\frac{r^2m_\gamma(r)}{r-3m_\gamma(r)}.
	\label{eq:circular_EL}
\end{eqnarray}
Therefore the marginal-stability condition $\dd L^2/\dd r=0$ has the useful
mass-function form, which is also the local circular-orbit criterion employed
in Refs.~\cite{Acharyya2024,Maeda2025},
\begin{equation}
	\mathcal Q_\gamma(r)
	\equiv r^2m_\gamma'(r)+r m_\gamma(r)-6[m_\gamma(r)]^2=0,
	\label{eq:ISCO_mass_form}
\end{equation}
where $m_\gamma'(r)=4\pi r^2\rho_\gamma(r)$.  This form makes clear that stability
inside the matter-supported region can depend on the Dehnen index through both
$m_\gamma(r)$ and $\rho_\gamma(r)$.  The equivalent metric form is
\begin{equation}
	\boxed{
		\mathcal S_\gamma(r)
		\equiv3A_\gamma(r)A_\gamma'(r)
		+rA_\gamma(r)A_\gamma''(r)
		-2r[A_\gamma'(r)]^2=0.}
	\label{eq:ISCO_general}
\end{equation}
Stable circular orbits lie on the side where $\mathcal Q_\gamma(r)>0$
(equivalently $\mathcal S_\gamma(r)>0$), subject also to
$2A_\gamma(r)-rA_\gamma'(r)>0$.  In the vacuum gap,
$A_\gamma(r)=C_\gamma(1-\rh/r)$, so $C_\gamma$ cancels from
Eq.~\eqref{eq:ISCO_general} and gives
$r_{\rm ISCO}=3\rh=6M_{\bullet,\gamma}$.
Consequently, if the halo is required to begin no deeper than the vacuum
Schwarzschild ISCO, one should impose
$\rin\geq r_{\rm ISCO}$, or equivalently $\kappa\geq3$.
For matter extending beyond the edge, Eq.~\eqref{eq:ISCO_general} should also
be checked through the halo because self-gravity can create additional stable
or unstable bands.  We therefore distinguish the effective anisotropic-fluid
model, for which $\kappa\geq\kappa_{\rm DEC,\gamma}$, from a literal stable
Einstein cluster, for which
$\kappa\geq\max\{\kappa_{\rm DEC,\gamma},3\}$ and
$\mathcal S_\gamma(r)\geq0$ throughout the occupied branch.
The latter requirement is intentionally more restrictive: satisfying the DEC
does not by itself prove microscopic orbital stability.

\subsubsection{Characteristic values of \texorpdfstring{$\kappa$}{kappa}}

The values $\kappa=3/2$ and $\kappa=3$ answer different questions.  Since the
vacuum photon sphere is $r_{\rm ph}=3\rh/2$ and $\rin=\kappa\rh$,
$\kappa=3/2$ only separates an orbit in the matter region
($\kappa<3/2$) from one in the vacuum gap ($\kappa>3/2$); the equality places
the orbit at the halo edge.  Section~\ref{sec:photon} derives these cases from
the null-geodesic equation.

The value $\kappa=3$, by contrast, follows from demanding that massive
particles at the halo edge begin at or outside the marginally stable orbit.
For the smooth depleted profile,
$m_\gamma(\rin)=\rh/2$ and $m_\gamma'(\rin)=0$, and hence
\begin{equation}
	\mathcal Q_\gamma(\rin)
	=\frac{\rh^2}{2}(\kappa-3).
	\label{eq:K_edge_kappa}
\end{equation}

Consequently, stability at the halo edge requires $\kappa\geq3$. This
bound is independent of $\gamma$ because the density and its contribution
to the mass gradient vanish at $\rin$, so the local edge condition reduces
to its Schwarzschild form. The corresponding dimensional radii nevertheless
remain profile dependent through the dressed horizon $\rh$.
The three characteristic bounds therefore have distinct physical meanings,
as summarized in Table~\ref{tab:kappa_bounds}.
\begin{table*}[t]
	\centering
	\caption{Characteristic constraints on the inner halo edge
		$\rin=\kappa\rh$. The DEC bound refers to local admissibility of the
		effective anisotropic source at the edge, while the latter two conditions
		concern the circular motion of the massive constituents. The
		$\kappa\geq3$ condition ensures stability at the edge; stability
		throughout the occupied halo must be checked separately.}
	\label{tab:kappa_bounds}
	\small
	\begin{tabular}{@{}lll@{}}
		\toprule
		Condition at the halo edge & Bound & Physical interpretation \\
		\midrule
		Dominant energy condition
		& $\kappa\geq5/4$
		& Effective anisotropic fluid \\
		Timelike circular motion
		& $\kappa>3/2$
		& Massive cluster particles \\
		Stable circular motion
		& $\kappa\geq3$
		& Stable-cluster edge benchmark \\
		\bottomrule
	\end{tabular}
\end{table*}
In the convention of Ref.~\cite{Shen2025}, $\nu_{\rm SWY}=2\kappa$, so the same hierarchy becomes $\nu_{\rm SWY}\geq5/2$, $\nu_{\rm SWY}>3$, and $\nu_{\rm SWY}\geq6$. The first of these reproduces the DEC bound obtained in Ref.~\cite{Shen2025}, while the latter two follow from the additional
requirements of timelike and stable circular motion of the cluster
constituents.

For the same Schwarzschild-normalized benchmark, $\Mh/\Mbare=10$ and $a/\Mbare=1000$, we numerically evaluate
$\mathcal Q_\gamma(r)$ in Eq.~\eqref{eq:ISCO_mass_form} throughout the occupied halo. For all four profiles, its global minimum occurs at the inner edge. The corresponding critical stability parameters and radii are summarized in Table~\ref{tab:stability_threshold}.
\begin{table}[t]
	\centering
	\caption{Critical stability parameters for the four depleted Dehnen
		profiles with $\Mh/\Mbare=10$ and $a/\Mbare=1000$. For each profile,
		the global minimum of $\mathcal Q_\gamma(r)$ occurs at the halo edge,
		giving the common threshold $\kappa_{\rm stab,\gamma}=3$. The last
		two columns show the corresponding dressed-horizon and halo-edge
		radii normalized by the reference Schwarzschild horizon.}
	\label{tab:stability_threshold}
	\small
	\begin{tabular}{@{}cccc@{}}
		\toprule
		$\gamma$
		& $\kappa_{\rm stab,\gamma}$
		& $\rh/\rhSch$
		& $\rin/\rhSch$
		\\
		\midrule
		$0$   & $3$ & $1.000002122$ & $3.000006365$ \\
		$1$   & $3$ & $1.000355971$ & $3.001067911$ \\
		$2$   & $3$ & $1.063399454$ & $3.190198361$ \\
		$5/2$ & $3$ & $2.120980327$ & $6.362940979$ \\
		\bottomrule
	\end{tabular}
\end{table}
As shown in Table~\ref{tab:stability_threshold}, no additional minimum of $\mathcal Q_\gamma(r)$ develops within the matter-supported region for this diffuse, halo-dominated benchmark. The global stability threshold therefore remains $\kappa_{\rm stab,\gamma}=3$ for all four profiles, although the corresponding dimensional radii retain their dependence on the cusp index through the dressed horizon.

Hence the stability threshold is the universal value $\kappa_{\rm stab,\gamma}=3$, whereas $\rh$, $\rin$, and the absorbed mass retain their genuine $\gamma$ dependence.  A profile-dependent
critical $\kappa$ could arise only in a sufficiently compact halo that develops an additional minimum of $\mathcal Q_\gamma(r)$ in the nonzero-density region, or under a different edge prescription for which the density does not vanish at $\rin$.

\begin{figure*}[!htbp]
	\centering
	\includegraphics[width=0.76\textwidth]{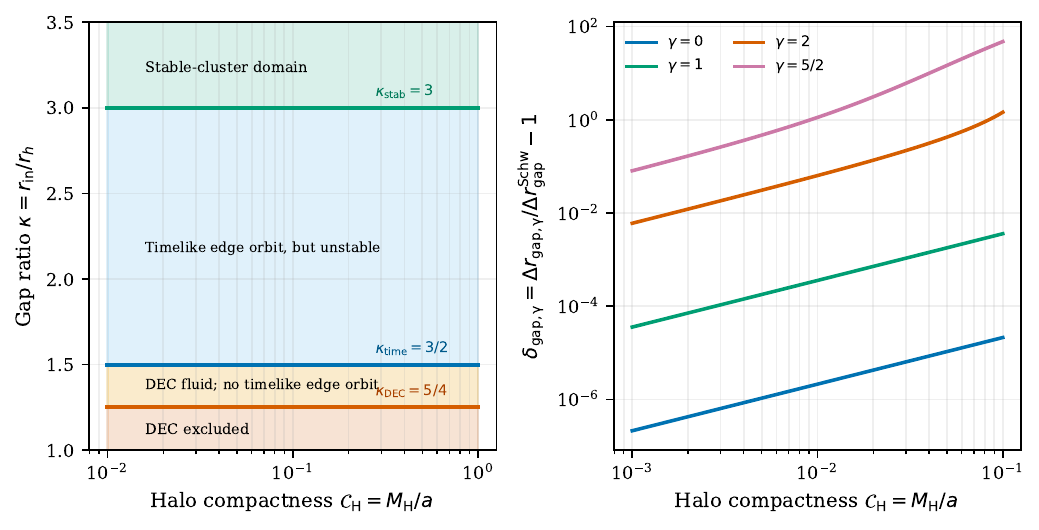}
	\caption{Edge admissibility and gap sensitivity.  The left panel separates
		the DEC-fluid, timelike-orbit, and stable-cluster domains.  The right panel
		shows the cusp-dependent fractional enhancement of the physical vacuum-gap
		width for $\kappa=3$ and $a/\Mbare=1000$.}
	\label{fig:edge_admissibility_map}
\end{figure*}

Figure~\ref{fig:edge_admissibility_map}
varies the gap parameter $\kappa$ at fixed $\Mh/\Mbare=10$ and tests the DEC,
timelike-orbit, and stable-orbit conditions, thereby summarizing the three
characteristic values of $\kappa$.  At fixed $\kappa$, the normalized gap width
is simply $(r_{\rm in,\gamma}-r_{h,\gamma})/r_{h,\gamma}=\kappa-1$ and has no
cusp dependence.  The physical width
$\Delta r_{\rm gap,\gamma}=(\kappa-1)r_{h,\gamma}$ remains $\gamma$ dependent
through the dressed horizon, as shown in the right panel.

The bounds established above apply to the diffuse galactic-halo regime studied in this work.  A sufficiently compact exterior distribution may develop additional circular-orbit branches or zeros of the radial metric function. Such ultracompact configurations lie outside the present astrophysical scope and require a separate global bifurcation analysis. The preceding test establishes the existence and radial stability of the individual circular geodesics underlying the mean-field cluster.  It is not a proof of collective Einstein--Vlasov stability under perturbations of the phase-space distribution and metric \cite{Andreasson2011,AndreassonRein2006}.  Ref.~\cite{Maeda2025} likewise identifies this collective problem as requiring a separate analysis. Establishing it for the depleted Dehnen family would require specifying a distribution function and evolving the coupled linearized Einstein--Vlasov system; we do not assume that stronger result here.  We now turn to the photon sphere and shadow observables of the admissible configurations.

%%%%%%%%%%%%%%%
%%%%%%%%%%%%%%%
%%%%%%%%%%%%%%%

\section{Photon sphere and shadow scale}
\label{sec:photon}

We now determine the circular null orbits and their critical impact parameters.
The analysis is necessarily piecewise: the Schwarzschild expression applies
when the orbit lies in the vacuum gap, whereas the full $\gamma$-dependent
redshift function is required when it lies in the halo.

\subsection{Photon spheres and null-orbit stability}

Photon propagation follows from the standard geodesic Lagrangian
\cite{Chandrasekhar1983,PerlickTsupko2022}
\begin{equation}
	2\mathcal L
	=-A_\gamma(r)\dot t^{\,2}
	+\frac{\dot r^{\,2}}{B_\gamma(r)}
	+r^2 \left(\dot\theta^{\,2}
	+\sin^2\theta\,\dot\phi^{\,2}\right),
	\quad \mathcal L=0 ,
	\label{eq:null_lagrangian}
\end{equation}
where a dot denotes differentiation with respect to an affine parameter and
the last equality selects null geodesics.  Spherical symmetry allows us to set
$\theta=\pi/2$ without loss of generality.  The conserved energy and angular momentum are
$E=A_\gamma(r)\dot t$ and $L=r^2\dot\phi$, respectively.
The null constraint then gives
\begin{equation}
	\dot r^{\,2}
	=\frac{B_\gamma(r)}{A_\gamma(r)}
	\left[E^2-L^2\frac{A_\gamma(r)}{r^2}\right].
	\label{eq:null_radial}
\end{equation}
A circular null orbit at $r=r_{\rm ph,\gamma}$ obeys $\dot r=0$ and
$\dd(\dot r^{\,2})/\dd r=0$.  Away from a zero of $B_\gamma(r)$, these two
conditions reduce to the standard compact photon-sphere equation
\cite{PerlickTsupko2022}
\begin{equation}
	r_{\rm ph,\gamma}A_\gamma'(r_{\rm ph,\gamma})
	-2A_\gamma(r_{\rm ph,\gamma})=0.
	\label{eq:photon_general}
\end{equation}
Thus every radial quantity in the final physical condition is evaluated at
$r_{\rm ph,\gamma}$.  Alternatively, one may define
$F_{\rm ph,\gamma}(r)=rA_\gamma'(r)-2A_\gamma(r)$ and say that
$r_{\rm ph,\gamma}$ is a root of $F_{\rm ph,\gamma}(r)=0$; only in this latter
root-finding notation is it appropriate to retain a free variable $r$.

For $\kappa>3/2$, the circular photon orbit lies inside the Schwarzschild
vacuum gap, where $A_\gamma(r)=C_\gamma(1-\rh/r)$.  Substitution into
Eq.~\eqref{eq:photon_general} gives
\begin{equation}
	C_\gamma\left[
	\frac{\rh}{r_{\rm ph,\gamma}}
	-2\left(1-\frac{\rh}{r_{\rm ph,\gamma}}\right)
	\right]=0.
	\label{eq:photon_gap_intermediate}
\end{equation}
Since $C_\gamma$ is finite and nonzero, it can be divided out, yielding $	r_{\rm ph,\gamma}=\frac{3}{2}\rh$.
For this orbit to lie within the vacuum gap, it must satisfy $r_{\rm ph,\gamma}<\rin=\kappa\rh$. Using the result above, this condition
reduces to $\kappa>\frac{3}{2}$.
%\begin{equation}
%	\kappa>\frac{3}{2}.
%	\label{eq:photon_gap_condition}
%\end{equation}
At the limiting value $\kappa=3/2$, the photon orbit coincides with the smoothly matched halo edge.

For $1<\kappa<3/2$, the photon orbit lies in the matter region and the complete
exterior lapse must be used.  From
$A_\gamma(r)=q_h(r)\exp[\Gamma_\gamma(r)]$ one obtains
\begin{equation}
	\frac{rA_\gamma'(r)}{A_\gamma(r)}-2
	=\frac{\rh}{r-\rh}+r\Gamma_\gamma'(r)-2.
	\label{eq:photon_phase_root}
\end{equation}
Differentiating Eq.~\eqref{eq:Gamma_general} gives
$\Gamma_\gamma'(r)=2h_\gamma(r)/
\bigl\{(r-\rh)[r-\rh-2h_\gamma(r)]\bigr\}$.
Evaluating every term at the circular orbit then gives the exact exterior
photon-sphere equation
\begin{widetext}
\begin{equation}
		\frac{\rh}{r_{\rm ph,\gamma}-\rh}
		+\frac{2r_{\rm ph,\gamma}h_\gamma(r_{\rm ph,\gamma})}
		{(r_{\rm ph,\gamma}-\rh)
			[r_{\rm ph,\gamma}-\rh-2h_\gamma(r_{\rm ph,\gamma})]}
		-2=0.
	\label{eq:photon_exterior_exact}
\end{equation}
\end{widetext}
Equation~\eqref{eq:photon_exterior_exact}, together with
$r_{\rm ph,\gamma}\geq\rin$, must be solved separately for each $\gamma$.
Unlike the vacuum-gap result $r_{\rm ph,\gamma}=3\rh/2$, it depends on the complete accumulated halo
mass $h_\gamma(r)$ and can admit more than one circular null orbit.  The physical
shadow boundary is associated with the relevant unstable orbit; its stability
is determined by the sign of
$\dd^2[A_\gamma(r)/r^2]/\dd r^2$ at that root.

There is a useful independent check on every numerical root.  Substituting the
Einstein-cluster equation
$A_\gamma'(r)/A_\gamma(r)=
2m_\gamma(r)/\{r[r-2m_\gamma(r)]\}$ directly into the circular-null
condition gives $(2m_\gamma(r_{\rm ph,\gamma}))/(r_{\rm ph,\gamma}-2m_\gamma(r_{\rm ph,\gamma})) = 2$ which leads to $r_{\rm ph,\gamma}=3m_\gamma(r_{\rm ph,\gamma})$.
%\begin{equation}
%	\frac{2m_\gamma(r_{\rm ph,\gamma})}
%	{r_{\rm ph,\gamma}-2m_\gamma(r_{\rm ph,\gamma})}=2,
%	\qquad\Longrightarrow\qquad
%	r_{\rm ph,\gamma}=3m_\gamma(r_{\rm ph,\gamma}).
%	\label{eq:photon_mass_identity}
%\end{equation}
Since $m_\gamma(r)=\rh/2+h_\gamma(r)$, this may also be written as
\begin{equation}
	\frac{r_{\rm ph,\gamma}}{\rh}
	=\frac32+\frac{3h_\gamma(r_{\rm ph,\gamma})}{\rh},
	\qquad
	B_\gamma(r_{\rm ph,\gamma})=\frac13.
	\label{eq:photon_normalized_identity}
\end{equation}
These identities show why the absolute radius and the radius normalized by
the dressed horizon need not have the same ordering with $\gamma$.

\begin{figure*}[!htbp]
	\centering
	\includegraphics[width=0.82\textwidth]{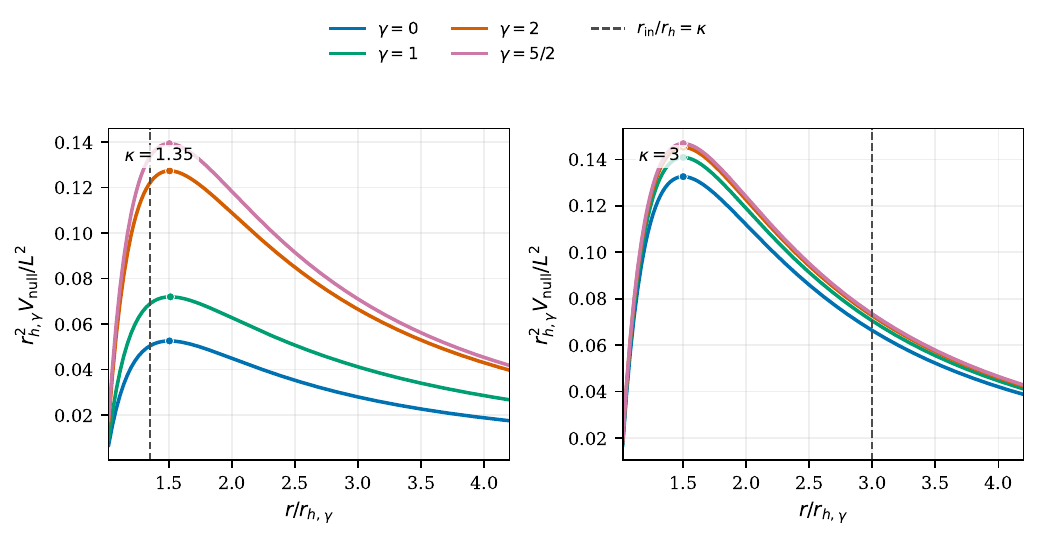}
	\caption{Dimensionless null effective potential for the four Dehnen models
		at $\Mh/\Mbare=10$ and $a/\Mbare=10$.  The orbit lies in the halo for
		$\kappa=1.35$ and in the vacuum gap for $\kappa=3$; the dashed line marks
		the corresponding halo edge.  The $\kappa=1.35$ curves represent the
		DEC-admissible effective-fluid branch, not a cluster of timelike particles
		at its inner edge.}
	\label{fig:baseline_null_potential}
\end{figure*}

The turning-point structure of the radial equation is governed by the null
effective potential $V_{{\rm null},\gamma}(r)=L^2A_\gamma(r)/r^2$.  Its extrema
obey Eq.~\eqref{eq:photon_general}; a local maximum is an unstable light ring,
whereas a local minimum is a stable one.  Figure~\ref{fig:baseline_null_potential}
compares the four discrete profiles when the principal orbit lies in the halo
and when it lies in the vacuum gap.  In the latter regime the maximum remains
at $r/r_{h,\gamma}=3/2$, although its height retains the exterior redshift
through the matching constant.

\subsection{Shadow and light-ring observables}

\begin{figure*}[t]
	\centering
	\includegraphics[width=0.96\textwidth]{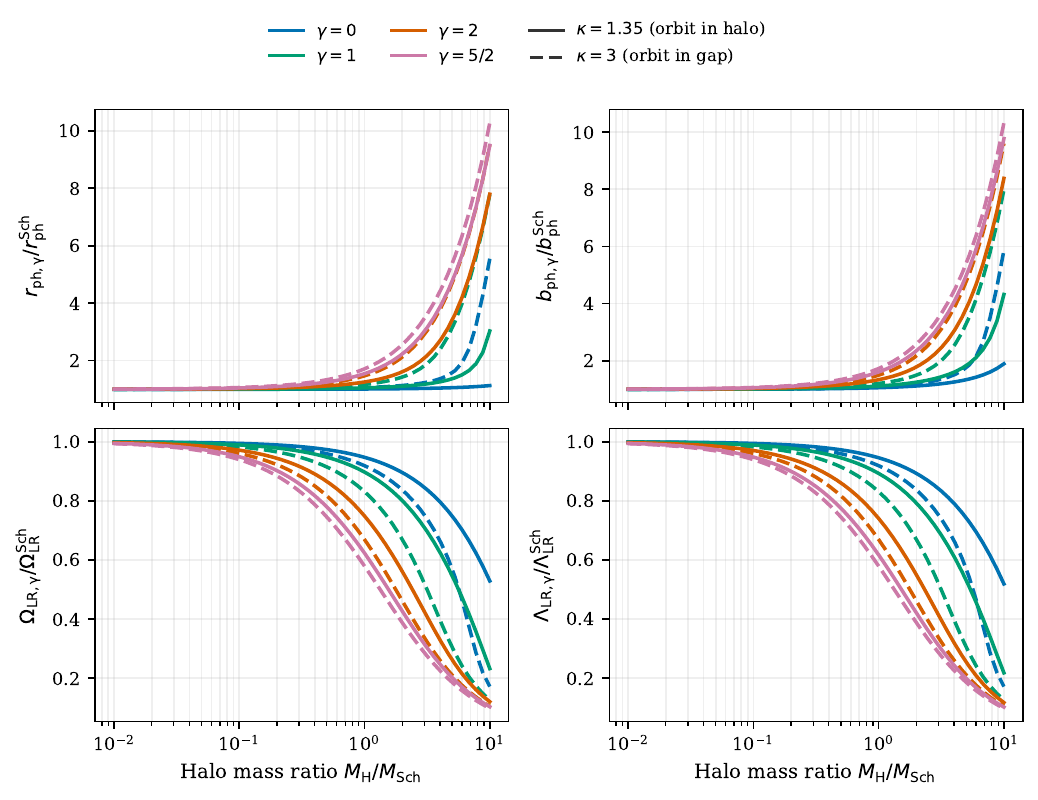}
	\caption{Schwarzschild-normalized photon-sphere, shadow, and light-ring
		diagnostics versus $\Mh/\Mbare$ for $a/\Mbare=10$.  The top row gives
		$r_{\rm ph,\gamma}/\rphSch$ and $b_{\rm ph,\gamma}/\bphSch$; the bottom row
		gives $\Omega_{\rm LR,\gamma}/\Omega_{\rm LR}^{\rm Sch}$ and
		$\Lambda_{\rm LR,\gamma}/\Lambda_{\rm LR}^{\rm Sch}$.  Color labels the four
		discrete Dehnen profiles.  Solid curves
		denote $\kappa=1.35$, for which the null orbit lies in the halo, and dashed
		curves denote $\kappa=3$, for which it lies in the vacuum gap.  All curves
		recover the Schwarzschild limit as $\Mh\to0$.  The comparison separates the
		change caused by the dressed horizon from the additional exterior-redshift
		contribution to the shadow through $C_\gamma$.  The solid curves are an
		effective-fluid diagnostic; the dashed curves satisfy the stable-cluster
		edge criterion.}
	\label{fig:observables}
\end{figure*}

Here $C_\gamma\equiv\exp[\Gamma_\gamma(\rin)]
=A_\gamma^{\rm ext}(\rin)/(1-\rh/\rin)>0$ 
is the matching constant at the halo edge. It relates the normalization of
the timelike Killing coordinate in the vacuum gap to the asymptotically
normalized time coordinate at infinity. For a distant observer, the critical
impact parameter associated with the photon sphere is
\cite{PerlickTsupko2022}
\begin{equation}
	b_{\rm ph,\gamma}
	=
	\frac{r_{\rm ph,\gamma}}
	{\sqrt{A_\gamma(r_{\rm ph,\gamma})}}.
\end{equation}
When $\kappa>3/2$, the photon sphere lies in the vacuum gap, where
$r_{\rm ph,\gamma}=3\rh/2$ and
$A_\gamma(r_{\rm ph,\gamma})=C_\gamma/3$. Hence
\begin{equation}
	b_{\rm ph,\gamma}
	=
	\frac{3\sqrt{3}}{2}
	\frac{\rh}{\sqrt{C_\gamma}}.
	\label{eq:shadow_general}
\end{equation}
Thus, in this regime, the coordinate radius of the photon sphere is set
entirely by the dressed horizon, whereas the shadow scale retains an
additional dependence on the exterior halo through the matching factor $C_\gamma$.
To compare different cusps without mixing physical dimensions, we introduce the ratios
\begin{equation}
	\mathcal R_{r,\gamma}=\frac{r_{\rm ph,\gamma}}{\rphSch},
	\qquad
	\mathcal R_{b,\gamma}=\frac{b_{\rm ph,\gamma}}{\bphSch}.
	\label{eq:schwarzschild_normalized_observables}
\end{equation}
These quantities are shown together with the corresponding frequency and Lyapunov ratios in Fig.~\ref{fig:observables}.  When $\kappa=1.35$, matter is present at the null
orbit and changes both its location and gravitational redshift.  When
$\kappa=3$, the orbit remains exactly $3\rh/2$ in the vacuum gap, but its
shift relative to the reference Schwarzschild value records the absorbed mass
through $\rh$; the impact parameter additionally records the exterior halo
through $C_\gamma$.  All four ratios approach unity as $\Mh\rightarrow0$,
providing a direct vacuum-limit check.

The numerical comparison across the halo edge uses the same exact equations,
without introducing a separate root function.  For the deliberately compact
Schwarzschild-normalized parameters $\Mh/\Mbare=10$ and $a/\Mbare=10$, the choice
$\kappa=1.35$ places the light ring in matter and makes its profile dependence
visible, whereas $\kappa=3$ places it in the vacuum gap and gives the common
normalized radius $r_{\rm ph,\gamma}/r_{h,\gamma}=3/2$.  The compact benchmark
is diagnostic rather than representative of the galactic hierarchy.  It
satisfies the DEC and has no additional exterior horizon for the four plotted
profiles, but because $\kappa<3/2$ it is not a literal timelike Einstein
cluster at the halo edge.  For
$\kappa=1.35$, the exact ratios $r_{h,\gamma}/\rhSch$ for
$\gamma=0,1,2,$ and $5/2$ are respectively
$1.126975$, $3.010222$, $7.772792$, and $9.479768$, while the corresponding
ratios $r_{\rm ph,\gamma}/\rphSch$ are $1.129141$, $3.030600$, $7.794040$,
and $9.491345$.  The physical radius increases with cusp index,
but its horizon-normalized value need not be monotonic because
Eq.~\eqref{eq:photon_normalized_identity} removes the dominant growth of the
dressed horizon.

The coordinate angular frequency and Lyapunov exponent supply complementary
dynamical diagnostics.  With the time coordinate normalized at infinity,
\begin{eqnarray}
\Omega_{\rm LR,\gamma}
		&=&\frac{\sqrt{A_\gamma(r_{\rm LR,\gamma})}}{r_{\rm LR,\gamma}},\\
\Lambda_{\rm LR,\gamma}^{2}
		&=&-\frac{B_\gamma(r_{\rm LR,\gamma})r_{\rm LR,\gamma}^{2}}{2}
		\left.\frac{\dd^2}{\dd r^2}\left(\frac{A_\gamma(r)}{r^2}\right)
		\right|_{r_{\rm LR,\gamma}}.
	\label{eq:LR_frequency_lyapunov}
\end{eqnarray}
Here the complete factor
$A_\gamma=(1-\rh/r)e^{\Gamma_\gamma}$ and its derivatives are retained.
Using the Einstein-cluster equation together with
$r_{\rm LR,\gamma}=3m_\gamma(r_{\rm LR,\gamma})$ reduces the second result to
\begin{eqnarray}
\Lambda_{\rm LR,\gamma}
		&=&\Omega_{\rm LR,\gamma}
		\sqrt{1-3m_\gamma'(r_{\rm LR,\gamma})},\\
\tau_{\rm LR,\gamma}&=&\Lambda_{\rm LR,\gamma}^{-1}.
	\label{eq:LR_lyapunov_mass}
\end{eqnarray}
Thus $\Lambda_{\rm LR}<\Omega_{\rm LR}$ when the orbit lies in nonzero-density
matter.  For $\kappa=5/4$ and the diffuse benchmark
$\Mh/\Mbare=10$ and $a/\Mbare=1000$, the nearly equal Schwarzschild-normalized
values for $\gamma=0$ differ only after seven decimal places:
$\Omega_{\rm LR}/\Omega_{\rm LR}^{\rm Sch}=0.9950013450$ and
$\Lambda_{\rm LR}/\Lambda_{\rm LR}^{\rm Sch}=0.9950012786$.  Their apparent equality at lower precision
is only rounding.  At $\kappa=3/2$ the orbit lies at the smooth zero-density
edge, and for $\kappa>3/2$ it lies in the vacuum gap; in both cases
$m_\gamma'(r_{\rm LR})=0$ and
\begin{equation}
\Lambda_{\rm LR,\gamma}=\Omega_{\rm LR,\gamma}
		=\frac{2\sqrt{C_\gamma}}{3\sqrt3\,r_{h,\gamma}}.
	\label{eq:LR_gap_rates}
\end{equation}

Applying the eikonal light-ring correspondence \cite{CardosoQNM2009} to the
exact rates obtained above gives the following result for the present depleted
Dehnen family:
\begin{widetext}
\begin{equation}
		\omega_{\ell n,\gamma}=
		\begin{cases}
			\displaystyle
			\frac{2\sqrt{C_\gamma}}{3\sqrt3\,r_{h,\gamma}}
			\left[\ell-\mathrm{i}\left(n+\frac12\right)\right],
			& r_{\rm LR,\gamma}\leq r_{\mathrm{in},\gamma},\\[3mm]
			\displaystyle
			\Omega_{\rm LR,\gamma}
			\left[\ell-\mathrm{i}\left(n+\frac12\right)
			\sqrt{1-3m_\gamma'(r_{\rm LR,\gamma})}\right],
			& r_{\rm LR,\gamma}>r_{\mathrm{in},\gamma},
		\end{cases}
		\quad +O(\ell^{-1}).
	\label{eq:eikonal_QNM}
\end{equation}
\end{widetext}
In the first branch, realized at the smooth edge for $\kappa=3/2$ and in the
vacuum gap for $\kappa>3/2$, the real and damping parts acquire the same
$\gamma$-dependent scale factor
$2\sqrt{C_\gamma}/(3\sqrt3\,r_{h,\gamma})$.
The cusp affects this branch through the absorbed mass in $r_{h,\gamma}$ and
through the exterior-halo redshift encoded in $C_\gamma$, even though the
light ring itself lies in vacuum.

The second branch applies when the light ring is inside the occupied halo.
Its real part continues to be fixed by the exact redshift function through
$\Omega_{\rm LR,\gamma}$, whereas its damping part contains the additional
local factor $\sqrt{1-3m_\gamma'(r_{\rm LR,\gamma})}$.  Since
$m_\gamma'=4\pi r^2\rho_\gamma>0$ there, the magnitude of the imaginary part
is reduced relative to the real-frequency scale.  The separation visible
between the solid curves in the lower panels of Fig.~\ref{fig:observables} is
therefore a direct signature of matter at the light ring, rather than merely
the global redshift of an exterior halo.  For the diffuse benchmark quoted
above this effect is extremely small, while the compact benchmark used in the
figure makes the profile dependence resolvable.  These statements concern the
eikonal test-field sector; coupled perturbations of the metric and cluster
would require a separate analysis.

%%%%%%%%%%%%%%%
%%%%%%%%%%%%%%%
%%%%%%%%%%%%%%%

\section{Conclusions}\label{sec:conclusion}

We have constructed a static final-state geometry that connects an initial
Dehnen halo to a dressed BH and a surviving exterior distribution
within one conserved mass budget.  The essential modeling input is explicit:
the initial mass inside the prescribed halo edge is assigned to the central
object.  The resulting horizon is therefore cusp dependent, whereas exterior
spherical matter alone does not shift a horizon whose final BH mass is
already fixed.  The region between this horizon and the remaining halo is a
vacuum gap, not a claim that DM can never cross the horizon.

The smoothly depleted profile yields continuous density and mass functions
and matches both metric functions at the halo edge without a thin shell.  The
general solution is asymptotically flat, has the correct conserved ADM mass,
and reduces to Schwarzschild when the halo is removed.  Its explicit
$\gamma=0,1,2,$ and $5/2$ members show that steeper initial cusps place more
mass near the center and consequently generate larger dressed horizons for
fixed initial data.

Physical admissibility separates three independent requirements.  The local
dominant-energy bound fixes the minimum edge ratio at $\kappa=5/4$, but the
associated edge speed is superluminal and this limiting configuration can only
be interpreted as an effective anisotropic fluid.  Timelike circular motion
requires $\kappa>3/2$, with equality representing the limiting null edge,
whereas $\kappa=3$ places the halo edge at the vacuum ISCO and supplies the
stable-cluster benchmark.  Global bounds can become profile dependent for
sufficiently compact halos, but such ultracompact configurations lie outside
the diffuse galactic-halo regime considered here.

The photon-sphere and shadow analysis further distinguishes dimensional from
horizon-normalized effects.  Absolute strong-field scales generally grow with
the cusp through the dressed mass, while normalization by the dressed horizon
can suppress this ordering.  When the photon sphere lies in the vacuum gap its
coordinate radius is fixed by the dressed horizon, but the shadow retains the
redshift contribution of the exterior halo through the matching constant.
The angular frequency and Lyapunov exponent retain the full exponential
redshift phase.  They coincide for a light ring at the smooth halo edge or in
the vacuum gap, while nonzero matter at the orbit lowers the Lyapunov exponent
relative to the angular frequency and lengthens the instability timescale.
In the eikonal test-field limit, these two quantities determine respectively
the oscillatory and damping parts of the quasinormal frequency, so a
matter-supported light ring increases the mode lifetime relative to its
oscillation period.

Earlier studies of DM spikes introduced an inner scale for purposes
different from the stationary Einstein-cluster boundary considered here.
Gondolo and Silk~\cite{GondoloSilk1999} employed a phenomenological
relativistic correction to an adiabatically generated spike, for which the
density was taken to vanish at $4R_{\rm S}$, where
$R_{\rm S}=2M_{\rm BH}$.  A fully relativistic phase-space calculation by
Sadeghian, Ferrer, and Will~\cite{Sadeghian2013} subsequently moved this
capture-induced zero to $2R_{\rm S}$.  Merritt \emph{et al.}%
~\cite{Merritt2002}, by contrast, showed that BH mergers and binary
heating can substantially erode or destroy a central spike, and therefore do
not select a universal sharp inner radius.  The radius
$r_{\mathrm{in},\gamma}=\kappa r_{h,\gamma}$ obtained in the present
construction has a different and complementary meaning: it is the inner
support boundary of the final stationary Einstein cluster around the dressed
horizon $r_{h,\gamma}=2M_{\bullet,\gamma}$.  The dominant energy condition
permits an effective-fluid boundary at $\kappa=5/4$, timelike circular
constituents require $\kappa>3/2$, and a cluster composed of stable circular
orbits requires the stronger benchmark $\kappa\geq3$ in the diffuse-halo
regime.  Consequently, the physically preferred stable-cluster edge,
$r_{\mathrm{in},\gamma}\geq3r_{h,\gamma}$, lies between the relativistic
capture scale $2R_{\rm S}$ and the earlier phenomenological value
$4R_{\rm S}$ when the radii are compared using the same central mass.
Unlike those fixed capture prescriptions, however, both
$r_{h,\gamma}$ and $r_{\mathrm{in},\gamma}$ retain information about the
initial Dehnen cusp through the absorbed mass.  Our result should therefore
not be interpreted as replacing the capture radii of the phase-space spike
models; rather, it supplies the additional orbital-consistency condition
required when the surviving matter is represented as a stationary Einstein
cluster.

The fixed-final-mass geometries of Refs.~\cite{Shen2025,Cardoso2022} arise as
controlled limits of the same integrated solution, while retaining their distinct
fixed-final-mass interpretation.  The present construction is an equilibrium
model rather than a dynamical capture calculation; determining the depletion
profile and accreted fraction from kinetic evolution is the natural next step
toward applications to BH imaging, stellar dynamics, and
extreme-mass-ratio inspirals.

%\acknowledgments
%acknowledgments

\bibliographystyle{mybibstyle}
\setlength{\emergencystretch}{3em}

\bibliography{bib}

\end{document}